\documentclass[master=mai,masteroption=ecs]{kulemt}
\usepackage{algorithm}
\usepackage{algpseudocode}
\usepackage{graphicx}  
\usepackage{float}  
\makeatletter
\let\c@lofdepth\relax
\let\c@lotdepth\relax
\makeatother
\usepackage{subfigure} 
\usepackage{url}
\usepackage{siunitx}
\usepackage{amssymb}
\usepackage{amsmath}

\usepackage{array}

\setup{
  title={Market Making via Reinforcement Learning in China Commodity Market},
  author={Jiang Junshu},
  promotor={Prof. Dr.Wim Schoutens},
  assessor={Jan De Spiegeleer \and Jente Van Belle},
  assistant={Thomas Dierckx}}

\setup{font=lm}

\usepackage[pdfusetitle,colorlinks,plainpages=false]{hyperref}

\begin{document}

\begin{preface}
I would like to thank everybody who kept me busy the last year, especially my promoter Prof. Dr. Wim Schoutens, and my mentor Thomas Dierckx who provided their expertise and support me. I would also like to thank the jury for reading the text. My sincere gratitude also goes to my girlfriend, Peggy Du, who encourages me and keeps me on track this academic year. Besides, I would like to thank Mr. Solaire Xiao from Honghui Investment$^\circledR$, Shenzhen, who provided support in maintaining the server where we deployed our Xtrader system and conducted the experiment, and also Mr. Feng Zijian from Nanjing University, who discussed with me upon some tricky mathematical formulas. 
\end{preface}

\tableofcontents*

\begin{abstract}
Market makers play an essential role in financial markets. A successful market maker should control inventory and adverse selection risks and provide liquidity to the market. As an important methodology in control problems, Reinforcement Learning enjoys the advantage of data-driven and less rigid assumptions, receiving great attention in the market-making field since 2018. However, although the China Commodity market has the biggest trading volume on agricultural products, nonferrous metals, and some other sectors, the study of applying RL to Market Making in China market is still rare. In this thesis, we try to fill the gap. Our contribution is threefold: We develop the Automatic Trading System and verify the feasibility of applying Reinforcement Learning in the China Commodity market. Also, we probe the agent's behavior by analyzing how it reacts to different environmental conditions. 
\\ \par\

\emph{\textbf{ Keywords \textemdash\ Market Making, Inventory Control, Adverse Selection, Reinforcement Learning, China Commodity Market, Automatic Trading System, Market Micro-structure Theory, Stochastic Control}}
\end{abstract}

\listoffiguresandtables
\chapter{List of Abbreviations and Symbols}
\section*{Abbreviations}
\begin{flushleft}
  \renewcommand{\arraystretch}{1.1}
  \begin{tabularx}{\textwidth}{@{}p{12mm}X@{}}
    ATS   & Automatic Trading System \\
    RL   & Reinforcement Learning \\
    MLP  & Multi-layer Perceptron \\
    SDE & Stochastic Partial Differential equation \\
    MABS & Multi-agent based Simulation \\
    MDP & Markov Decision Process \\
    MM & Market Maker \\
    DQN & Deep Q Network \\
    DNN & Deep Neural Network \\
    SHFE & Shanghai Future Exchange \\
    HJB & Hamilton-Jacobi-Bellman Equation \\
    MO & Market Order \\
    LO & Limit Order \\
    HFT & High-frequency Trading \\
    PDP & Partial Dependence Plot
  \end{tabularx}
\end{flushleft}

\section*{Symbols}
\begin{flushleft}
  \renewcommand{\arraystretch}{1.1}
  \begin{tabularx}{\textwidth}{@{}p{12mm}X@{}}
    $\alpha$    & learning rate  \\
    $\mathcal{A},\ S$   & Action Space, State Space\\

  \end{tabularx}
\end{flushleft}

\mainmatter

\chapter{Introduction}
\label{cha:intro}


Market making is a fast-growing business in financial markets, and it benefits the market by fostering liquidity\cite{elainewah2020mmWealfh}. Market making is one of the high-frequency tradings and is sensitive to exchanges' rules. Rules like transparency setting or data disclosure will significantly affect the performance of marking making. China's commodity future is a thriving market where the total trading amount is 87 trillion for 2021. Compared with a developed market like the Chicago Mercantile Exchange(CME), where the High-frequency trading(HFT) counts for 70\% of its daily trading volume, market-making in China is estimated to be around 30\%\cite{Huang2019ats}. Therefore, the percentage of HFT will continuously increase further in the foreseeable future. Traditionally method in market making focuses on treating it as a stochastic control problem and solving the corresponding stochastic Partial differential equation(SPDE) to obtain the optimal strategy. However, SPDE methods rely heavily on assumptions and cannot largely leverage the historical data. Besides, it is complicated to incorporate some constraints imposed by exchanges. Surprised by the gigantic success of Reinforcement Learning(RL) in control problems, an increasing number of studies in the market making has converted to applying RL since 2018. However, there are few studies on China's commodity market.

This study focuses on applying Reinforcement learning to market-making in China Commodity Future Market. Because of different regulation rules, market-making strategies developed in other markets cannot be directly deployed in China. We adapt the RL algorithm to China Commodity Future Market and develop the Automatic Trading System(ATS) for testing and deployment in the future. Specifically, our contributions are: 1) we design an RL agent and verify its effectiveness through a comparison experiment with a baseline model; 2) develop our own Xtrader ATS, which provides a convenient programming interface and supports multi-strategies; 3)conduct partial dependence analysis to get more insight of agent's behavior pattern.


\chapter{Background and Related Work}
\label{cha:back}

\section{Automated Trading System}
\label{cha:back:ats}
An automatic Trading System(ATS) is an IT system that can automatically submit electronic orders given by algorithms. The earliest ATS can date back to the last centure\cite{grossman1988} which firstly uses electronic order. Before the advent of ATS, the quote is manually submitted by human agents. In \cite{brogaard2010}, a typical high-frequency trading strategy conducted by human traders demonstrates an average holding time of fewer than 30 seconds. Nowadays, The reaction time of an HFT strategy has been pushed to the microsecond, even nanosecond level with an increasingly faster computer. Undoubtedly speed is one of the key factors in HFT. 

From the general picture, two promising directions of ATS developments are 1) integration of heterogeneous data types and 2)low-latency trading system. Traditional ATS mainly takes the historical quote data as the sole input. However, with increasingly fierce competition in the high-frequency regime, companies are searching for different types of data sources to gain their niche in the competition. \cite{gastli2021satelline} uses satellite images to facilitate stock price prediction. \cite{schumaker2009textual} uses breaking news as information sources besides historical price to predict stock price, \cite{chen2018incorporating} uses a knowledge graph for depicting the supply chain relationship and trend prediction, and \cite{dierckx2020using} uses Google News statistics to predict volatility. All the kinds of literature demonstrate that information sources for modern trading strategies would be increasingly diversified and complicated. 

Another tendency is the faster and faster low-latency system. The ATS for high-frequency trading is at least as important as the trading strategy itself. Because the fastest traders can seize a significantly large share of profit(it depends on more relative speed instead of absolute speed), companies are keen to invest millions of dollars to gain even one millisecond improvements. It can be exemplified by the famous story of the Spread Networks company, which spent 300\$ to construct a new high-speed cable connecting Chicago and New York's financial markets to improve the latency from 16 milliseconds to 13 milliseconds. However, this plan is soon bypassed by microwave technology which reaches 8.1 milliseconds latency\cite{budish2015high}. The crazy arms race in low-latency technology proves the importance of speed in high-frequency trading.

The latency of the whole trading can be further divided into two parts: external latency and internal latency. They refer to the time consumed on web transmission and the consumed by ATS's calculation. For rudimentary ATS, the external latency is the largest part. One can easily improve by renting a co-location host to deploy the ATS. The optimization of internal latency is more effort-taking. \cite{subramoni2010streaming} uses a customized network interface controller(NIC), which allows OS-bypass data communication and improves the latency from a data transmission perspective. \cite{tang2016fpga} optimize the speed from the code execution perspective. They use the Field Programmable Gate Array(FPGA) with logical codes and reach around 1us per market data, which is dozens of times faster than the system implemented in C language. Also, the logic of the strategy is designed as simply as possible to squeeze the execution time furthermore. \cite{Huang2019ats} points out it is not astonishing that high-frequency strategies are inclined to use a simple algorithm like linear regression.



\section{Market Micro-structure Theory}
\label{cha:back:ms}
Market Microstructure is a subfield of financial economics. Its main topics are about trading infrastructure and its relationship with price formation, market liquidity and etc\cite{garman1976market}. Because of transaction cost and two price setting of ask-bid, some finding in the high-frequency world is not fully aligned with their counterparts in the low-frequency world. Empirical studies find the market in the high-frequency setting is not fully efficient, and it is possible to predict the short tendency of price movement and volatility level\cite{figueroa2011estimation}. Besides, even in a long-term setting, price change still perverse some correlation\cite{kercheval2011risk}.

Market type can be divided into dealer market, order-driven market, and conglomerate market like specialist mechanism in the New York Stock Exchange(NYSE). In the dealer and specialist markets, market makers are explicitly designated by the exchange. Usually, dealers are obliged to foster the exchange's liquidity and reduce the temporary disparity between buyers and sellers. In return, makers may enjoy a more attractive commission rate and rebate.
Since trading activities are strongly sensitive to trading rules, regulators can design the rules to maximize the whole welfare of participators and trading volume. The trading rule of exchange has many aspects, including ante transparency, post transparency, market segmentation, commission, rebate, order types, and canceling limitations. Modeling and incorporating those complicated trading rules into trading strategy is essential to market-making. As for trading rule and data disclosure for China Commodity, we elaborate it in Chapter \ref{cha:meth}.

The risks and the potential rewards in the dealer market have been well studied. The potential reward is the spread, and in the ideal situation, the limit orders posed by the dealer on two sides are executed simultaneously, and the spread can be earned. However, this profit is never guaranteed due to several risks \textemdash inventory risk, adverse selection risk, execution risk, and latency risk. As for execution risk, it is rooted in the uncertainty essence of limit order. Market participators are less patient and are more inclined to execute immediately by market order when the market is more likely to stop, or the order is less like to be executed\cite{foucault1999order}.

Inventory risk is the risk of accumulating a net position in one direction and describes the potential loss if the price moves the opposite way. Stoll\cite{stoll1978supply} first studied the inventory risk and built an inventory model on a two-period setting and later be extended to multistep periods by Ho\cite{ho1981optimal}. The theoretical model built from those two studies states the dealer would quote more aggressively on the net position side by posing a more attractive limit order or even posing a market order to liquidate the unfavorable position eagerly. Therefore, it suggests the direction of trade series, inventory process of dealers, and quoted price should demonstrate mean-reverting patterns. Hansch conducted an empirical study with historical data from London Stock Exchange(LSE)  to verify those theoretical models \cite{hansch1998inv}. He found there is inventory reversion in inventory, and the strength of the reversion increases in the inventory divergence, which aligns with the inventory model. 

Adverse selection risk states dealer suffers from loss when traded with informed traders who know better about the underlying assets' price. In adverse selection theory, the participants in the market can be categorized as informed traders, dealers, and uninformed traders (also known as noise traders or liquidity traders). The classical model of adverse selection is introduced in \cite{glosten1985bid}. It assumes that the dealer can learn from the orderflow and update its belief on how likely this trade is launched by an informed trader. With the existence of informed traders, the spread will be larger than zero even when other risks and costs are absent. Because of the anonymity of trades, the dealer cannot tell informed traders from the others; therefore dealer can only indiscriminately increase the spread when he discerns there might be informed traders. Franck et al. introduce the concept of Percentage of Informed Trader(PIN)\cite{easley1992adverse} and the method to estimate it. Its high-frequency version of PIN is introduced in \cite{easley2012flow}. Adverse selection can also be studied from an informed traders' perspective. How to utilize the private information for an informed trader to maximize its profit
is studied in \cite{kyle1985continuous}. Gao introduces the latency risk\cite{gao2020optimalLatency}. With low-latency ATS, the agent may suffer less from adverse selection risks and obtain a better position in the waiting queue of the orderbook.

After realizing the existence of inventory risk, adverse selection risk, and execution risk, a natural question is: Can one estimate the percentage of each of those risk sources counts in the spread? This question is addressed in \cite{huang1997components}. Huang and Stoll introduce an econometrics method to estimate the percentage of each of those three risks in the spread.

Unlike the dealer market, the order-driven market is driven by orders. One can freely decide to submit limited orders or market orders to be a maker or dealer. Makers quote limit order(LO) and provide liquidity while takers submit market order(MO) and consume liquidity. The strategic choice between those two types of orders in an order-driven market was studied by Parlour\cite{parlour1998price}. It revealed that the status of the orderbook will affect the selection of order types. Specifically, a longer queue on the self side will make posing a limit order less attractive, and a longer queue on the opposite side will make it more appealing. This provides theoretical support for the effectiveness of the orderbook imbalance indicator in short trend prediction.

\section{Market Making via Stochastic Control}
\label{cha:back:mmsc}
Stochastic control is widely used in quantitative finance and is a popular methodology in market-making problems. The most classical problem in this field might be the Merton problem, in which the agent needs to allocate his wealth between risk and risk-free assets to maximize his terminal wealth\cite{merton1975optimum}. By formulating the SPDEs and the terminal return function, the origin problem can be converted to Hamilton-Jacobi-Bellman(HJB) equations, which is essentially a non-linear partial differential equation. Then, the optimal strategy solution can be obtained by solving the HJBs. Merton's problem is relatively simple since there is no interaction between the environment and agents. The price processes for the risky and risk-free assets are independent of the agent' strategy, and the agent's behavior would not change the environment. However, this is not the case in market making. In market making, the limit order posed on the orderbook will affect the length of the queue and the following execution probability.


The earliest stochastic control method in the market making can date back to the 1980s\cite{ho1981optimal} when Ho first formulate market making under the stochastic control framework. They assume the indifference price of these assets to the dealer subject to a diffusion process and the arrival of market order subject to a Poisson process whose sensitive($\lambda$) are affected by the order posting by the market maker. During the trading session, agents try to maximize their expected utility function by strategically posing the limit order. Avellaneda\cite{avellaneda2008high} modifies the model based on Ho's work and makes it adapt to the order-driven market. He assumes the mid-price instead the indifference price subject to the diffusion process and solves the HJB with analytical methods. Lately, Stoikov\cite{stoikov2009option} extends this idea to the problem of market-making in the option market. However, the methods above only tackle the inventory risk and ignore the adverse selection risks. Cartea incorporates short-term alpha by assuming the mid-price as a diffusion process with short-term drift\cite{cartea2018alpha}. By comparing the performance of two types of agents, the one without perceiving short-term alpha will be driven out by the one including short-term alpha.

\section{Market Making via Reinforcement Learning}
\label{cha:back:mmrl}
Reinforcement learning(RL) is an important method for control problems. The difference between the control problem and prediction problem can be characterized by the following two points: 1) in the control problem, the choice of action is also affected by the agent's state instead of purely by the environment 2) the agent interacts with the environment. By choosing different actions, the whole system involves in different directions. Both RL and Markov Decision Process(MDP) shares some same elements \textemdash the action space $\mathcal{A}$ and state-space $S$. The difference between RL and MDPs is that the probabilistic transition function and reward function are not available for RL. The general framework to solve MDPs and RLs is Bellman Equations. Bellman Equations bridge the value of two states with discount rewards between them. When the system model is available and the policy is given, the MDP degenerates into a linear system and can be directly solved. However, the computation is usually high because of the matrix's inverse operation. A more popular method is solving MDP by iterations.

MDPs can be solved by policy iteration and value iteration. The convergence property for them is proved in\cite{morton1977discounting}. One can get a table indicating the value of state-action pairs by solving MDP. This table can guide the agent on how to act, who only needs to look up the table and chooses the action with the highest value. Traditional RL can be categorized by dynamic programming, the Monte Carlo method, and temporal difference. Given the start state, In the Monte Carlo method, the samples will be drawn continuously until the path reaches the end state. In time different($TD(n)$), the samples will draw $n$ steps and will update the Q table before the next sampling. The Monte Carlo method and Time Difference have their pros and cons. Monte Carlo has a high variance since it has a longer sampling path. The time difference method will introduce bootstrapping bias because the model itself will affect the subsequent sampling path. One can view Monte Carlo as time different with an infinite number of time steps (until the game is over). A lot of variants of the RL model strive to combine the features from both methods to trade-off between high variance and bias. \cite{van2016deep} introduce RL model consists of the value function and action function to mitigate the bootstrapping bias. The action function is used to select action, while the update of weight will be applied to the value function on a regular basis. Another direction is focusing study multistep temporal differences and how to assign different weights to steps. The prioritized experience replay\cite{schaul2016prioritized} uses $TD(n)$ and gives weight to the reward series by calculating the importance-sampling weight.

Traditional RL cannot handle the scenario when state space is huge or is continuous. A lot of function approximation methods are introduced to address the curse of dimension. Function approximator can be categorized as linear method like using Fourier basis function\cite{konidaris2011value} and non-linear method like use kernel function\cite{connell1987learning}. Function approximator can also be categorized as parameter models or non-parameter models based on whether it incorporates prior knowledge. However, those methods still lack expressiveness when the state and action space are continuous and huge. Because of deep neural network(DNN)'s good expressiveness, DNNs are widely used as function approximators in RLs. The most famous application might be DQN\cite{mnih2015human}, which is also the model we adopted in this study. 

Since 2018, many studies have attempted to apply RL in market-making tasks. Spooner is the pioneer who first applied RL in market making\cite{spooner2018}. In this study, the agent is trained on simulation data, and tie coding is used as a function approximator. \cite{ganesh2019reinforcement} builds a simulation environment with different types of agents and trains an RL agent under this simulation environment. Agents trained in this competitive environment demonstrate more robust behavior. \cite{gueant2019deep} applying RL in a joint market-making task for multi assets. It is hard for SPDE and finite difference methods to solve multi-asset market-making tasks in high-dimensional space because of the curse of dimension. Spooner uses adversarial reinforcement learning to improve the robustness of the agent\cite{spooner2020robust}. In this paper, the whole system comprises a market maker agent and an adversary. The market maker's mission is to maximize his profit, while the adversary's mission is to select suitable parameters of the environment to minimize the market maker's profit. And those two models' parameters are trained interleaved. Although this study did not prove that the saddle point can be guaranteed to find, this paper demonstrates the robustness of the agent trained in an adversary setting is significantly higher than its counterpart in the vanilla setting. Zhong builds a pure data-driven market maker agent on the American stock market from the historical data\cite{zhong2020}. She carefully designs the state's space and trains the model on the historical data. Gasperov designs a compounded market maker agent which comprises two sub-systems\cite{gavsperov2021}. The signal generating unit is designed to generate predictive signals while the control unit conducts control action by taking the predictive signals and other states into consideration. This methodology can be paralleled as multi-task learning\cite{thung2018brief} in artificial intelligence. Chen builds the reinforcement agent when the exchange will return the rebates\cite{zhang2020reinforcement}. This demonstrates RL is helpful when the environment assumptions are hard to be formulated in SPDEs. The reinforcement learning technique builds a better control agent than the analytic approach in MM\cite{gavsperov2021review}.

Another promising application of RL in Market Making is Multi-agent based Simulation(MABS) of orderbook and orderflow. Intuitively, the behavior of the market is the result of interaction among a bunch of participates. Karpe builds a more realistic market simulation based on DRL of multi-agents\cite{karpe2020multi}. Experiments show the simulated market demonstrates some realistic features of orderflow. In reality, volumes of summited order in an equal length interval are more close to a gamma distribution, and the order arriving times are more close to Weibull distribution\cite{abergel2016limitOB}. Those features are often omitted when applying stochastic control methods in the market making, while those two features are observed in the MABS environments. Other works about market simulation can be found in \cite{li2020generating, byrd2019abides, vyetrenko2020get}.

\chapter{Preliminaries}
\label{cha:pre}

\section{Deep Reinforcement Learning}
\label{cha:pre:DRL}
In this study, the DQN model is used to build market-making agents. DQN is the deep version of the Q learning algorithm, a traditional algorithm in the Reinforcement learning sphere. In this section, we first introduce Markov Decision Process and reinforcement learning. Then, we elaborate on the DQN method used in this study.

\subsection{Markov Decision Process}
\label{cha:pre:DRL:MDP}
Markov Decision Process is the mathematical model of the sequential decision problem and can be used to solve the control problem when the system evolution has Markov property. Markov process can be defined by a quadruple $(S,\mathcal{A} ,P,R)$. $S$ refers to the state-space the Markov process $S_t$ can reach. $\mathcal{A}$ refers to the action space the agent can take at each time step. $P$ is the probabilistic transition function $P(S_t,a,\cdot)$ which stores the probabilistic distribution of the next state $S_{t+1}$ when action $a$ is taken at state $S_t$. When the state space is discrete, it can be represented by a square matrix with size $S\times A$. $R(S_t,a,S_{t+1})$ stands for the rewards from the environments. Therefore, the expected reward of taking action $a'$ when the state is $s$ can rewrite as $E_p[R(S_t,a')]=P_{S_t,a'}R(S_t,a',\cdot)$ function. Markov process refers to the stationary stochastic process, and the future state is irrelevant when the current state is given. It can be expressed by formula(strictly, this is a first-order Markov process):

\begin{equation}
P(S_{t+1}|S_t,S_{t-1} \cdots S_{t-n})=P(S_{t+1}|S_t)
\end{equation}
When the transition matrix is given, this property is implicitly satisfied because the distribution of the next state can be solely determined by the tuple $(S_t,a)$ regardless of previous states. Markov Decision Process can be solved by Policy iteration or Value iteration. Their Bellman equations are given in \ref{eq:valueIter} \& \ref{eq:policyIter} respectively. Both methods can be guaranteed to converge to a fixed point. The computational complexity for value and policy iteration is $O(SA^2)$ and $O(SA)$ respectively.

\begin{equation}
V_{k+1}(s)\leftarrow \max_a\sum_{s'}P(s,a,s')[R(s,a,s')+\gamma V_k(s')]
\label{eq:valueIter}
\end{equation}

\begin{equation}
V_{k+1}^\pi(s)\leftarrow \sum_{s'}P(s,a,s')[R(s,a,s')+\gamma V_k^\pi(s')]
\label{eq:policyIter}
\end{equation}

In equation \ref{eq:policyIter}, Policy $\pi: S\rightarrow A $ is a function  which map state space to action space. Therefore, Value iteration can be viewed as a special policy iteration with a greedy policy that always chooses the action to maximize the value function. The policy iteration can be further divided into policy evaluation and policy improvement steps. In the policy evolution, we solve the $V$ function given the fixed policy $\pi_i$, and in policy improvement, we optimize the policy from $\pi_i$ to $\pi_{i+1}$. Those two processes are conducted interleaved.

The policy evaluation can either be done by numerical methods or analytical methods. The equation in \ref{eq:policyIter} can be rewritten as a linear system $V=R+PV\gamma$ and the fixed point can also be solved directly by $V=(I-\gamma P)^{-1}R$ instead using numerical methods. However, the computational complexity is $O(S^3)$ in state-space size, which is unbearably high in practice. Therefore, numerical methods by iteration approximation are preferred when the state size is big.




\subsection{Reinforcement Learning}
\label{cha:pre:DRL:RL}
When the system model is unavailable or expensive to build, the MDPs need to be solved by reinforcement learning, a model-free method. The agent has no prior knowledge about the environment at first, and it is gradually built from the interaction with the environment. Because of the absence of system model, the value iteration and policy iteration should be re-wrote as \ref{eq:policyIterRL} and \ref{eq:valueIterRL}, which are the core formula used in the Q-Learning\cite{watkins1989learning} and SARSA\cite{rummery1994line}. 

\begin{equation}
    Q_{t+1}(S_t,a_t)\leftarrow Q_{t}(S_t,a_t)+\alpha_t[R_{t+1}+\gamma \max_{a_{t+1}}Q_{t}(S_{t+1},a_{t+1})-Q_{t}(S_t,a_t)]
    \label{eq:valueIterRL} 
\end{equation}

\begin{equation}
    Q_{t+1}(S_t,a_t)\leftarrow Q_{t}(S_t,a_t)+\alpha_t[R_{t+1}+\gamma Q_{t}(S_{t+1},a_{t+1})-Q_{t}(S_t,a_t)]
    \label{eq:policyIterRL} 
\end{equation}

The Q-Learning will converge to optimal $Q^*$ convergence in probability\cite{watkins1992q} when the following two conditions are satisfied:
\begin{itemize}
    \item $\sum_{i=0}^\infty{\alpha_i}=\infty$ and the limit of $\sum_{i=0}^\infty{{\alpha_i}^2}$ exists
    \item all pairs of $(S,A)$ is accessible
\end{itemize}
$\alpha$ refers to learning rate, and $\epsilon$ refers to explore rate. One typical series satisfies the first condition is $\alpha_t:=\frac{1}{t}$. The second condition can be satisfied with the $\epsilon-greedy$ search schema, which can balance between exploit and explore. One subtle difference between Q-learning and Sarsa algorithm is the former is off-policy, and Sarsa is on-policy. The concept of on/off-policy depends on whether the policy used to generate the samples(action function) is the target policy(value function). If both are the same, it is on-policy reinforcement learning; otherwise, it is off-policy reinforcement learning.

Another point that should bear into mind when developing the reinforcement algorithm is the design of the reward function. The agent's performance is strongly related to the design of the reward function. For example, if the reward is only given in the very last step for an episode game, the converging process at an early stage would be very slow. Besides, relevant domain knowledge and the expectation of the agent's behaviors should be encoded into the reward function when designing the RL algorithm. For our case in market making, to help our agent achieve some expected behavior like inventory control, we also carefully design the reward function \ref{cha:meth:env}. Therefore, the reward function is essential to the convergence of the model and is vital to getting the expected behavior pattern.


\subsection{Deep Q Network}
\label{cha:pre:DRL:DQN}
Methods introduced in subsection \ref{cha:pre:DRL:RL} can only be applied when the state space is discrete and of small size, and the value of each state-action pair can be stored in a two-dimension square matrix. However, When an agent goes to a large continuous state space, function approximation is necessary to make this method applicable in practice. In this study, we use Deep Q Network introduced in \cite{mnih2015human} which is the Q-learning algorithm equipped with a deep neural network to improve its expressiveness. Also, compared with Q-learning, experience replay and decoupling of value function are applied to make the training process more stable in DQN. Experience replay is used to reduce the correlation between samples. In a lot of scenarios, samples drawn over time are highly correlated. For example, in the racing competition, the location of the sports cars is continuous. It is the same problem when applied to the financial market, where the price and volatility are changing gradually. Strongly correlated samples are thought to contribute to the unstable training phase partly. Experience replay is introduced to make the samples more balanced over the sample space. Specifically, instead of updating for every step, experience replay will first accumulate a long episode steps and then randomly draw a batch of samples from those samples to update deep neural networks. By randomly sampling from a relatively large sample set, the correlation among samples can be reduced. In DQN, models are decoupled into the target model and value model. Target models are responsible for action selection, and the value model will store the updated weights. The target model's weights will be replaced by that of the value models regularly. Although both of those two methods introduced above cannot perfectly solve the problem of unstable convergence, they can at least mitigate this problem. Its pseudocode is given in Algorithm \ref{algo:dqn} to help demonstrate the idea of this algorithm.

\begin{algorithm}
    \caption{Deep Q-learning with Experience Replay and Action-Value Decouple}
    \label{algo:dqn}
    \begin{algorithmic}
    \State Initial buffer $D$, learning rate $\alpha$, explore rate $\epsilon$, soft update rate $\tau$, Environment Simulator $\mathbb{E}$
    \State Initial $Q$ network and copy into $A$ network
    \State $episodes \gets split\_suffle(dataset)$
    \For{$episode\ in\ episodes$} 

        \State with probability $\epsilon$ random select $a$, else $a=arg\max_{a'}{A(S,a')}$
    
        \State execute $a$ in $\mathbb{E}$ and observe $r,s'$
        \State push $(s,a,s',r)$back into buffer D

        \If{$D$ is full}
            \State batch $\gets (s,a,s',r)$ from D 
            \State $loss\gets Q(s,a)-[r+\gamma\times \max_a'Q(s',a')]$ 
            \State apply gradient descend and update into Q with learning rate $\alpha$
            \State $A\gets(1-\tau)+\tau Q$
            \State empty buffer $D$

        \EndIf
    \State $s\gets s'$
    \State $\alpha\gets \frac{1}{t}$

    \EndFor
    
    \end{algorithmic}
\end{algorithm}

In the origin study\cite{mnih2015human}, the function approximator is a convolutional neural network because its input is the pixels of the scene in Atari games. We use 4-layers Multi-layer Perceptron(MLP) as the function approximator mapping the state to a Q value vector of action-space size. Each component will represent the Q value of taking action $a$ at the state $s$. It is a regression task and the loss function is Mean Square Error(MSE) calculated by $Q(s,a)-[r+\gamma\times \max_a'Q(s',a')]$. Then standard back propagation algorithm can be used to calculate the gradient and optimize the parameters in the MLP. 


\section{Market Making}
\label{cha:pre:MM}


The discussion in this section is confined to high-frequency market-making in an order-driven market. Market maker, also known as a liquidity provider, plays an essential role in the financial market. Market maker quotes limit order on both sides in the market and provides liquidity for other participants, and ideally, will earn the spread as profit. Simply because there is no free launch in the financial market, the spread can also be viewed as compensation for all potential risks faced by the maker. 

Attitudes toward marking making are controversial. On the one hand, the Market maker can promote the overall wealfare\cite{elainewah2020mmWealfh,pithyachariyakul1986exchange}. With the competition among makers, the market's spread can be narrowed, and other participators can buy/sell the assets at less cost. On the other hand, HFT is criticized for attributing to flash crashes in financial markets. The volatility might be enlarged when the market is crowding bunches of agents with heterogeneous high-frequency strategies. 

The rest of the section is organized as follows: In this section, 
First, we explain concepts related to market-making, including different types of order and various types of action in subsection \ref{cha:pre:MM:ob}. Then, we explain market making under a stochastic control framework. \textbf{SPDE is still powerful and wide-used in market-making. However, since it is not the center of the discussion of this study, we only elaborate on the general framework of market-making via stochastic control in section \ref{cha:pre:MM:type}.}


\subsection{Orderbook and order types}
\label{cha:pre:MM:ob}
\begin{figure}
    \centering   
    \includegraphics[]{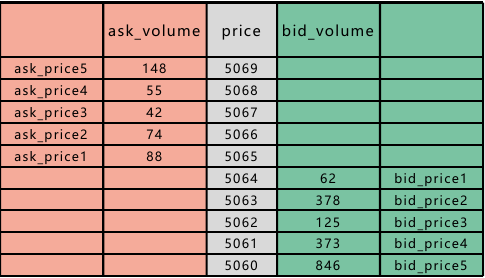}
    \caption{An example of orderbook}  
    \label{fig:ob}  
\end{figure}

In an order-driven market, orderbook are used to depict the current market environment and is a core concept in the market microstructure of order-driven market. Figure \ref{fig:ob} represent an example of an orderbook snapshot. The upper part is the ask side, where gather all the limit sell orders, and the lower part is the bid side, where gather all limit buy orders. Those orders are sorted by their distance to mid-price. Order closer to the center will have a higher priority in matching. The prices closest to the middle are called the best ask/bid price, which is the ask price1 5065 and the bid price1 5064 in this figure. They are the best price other participants can get if they decide to buy/sell a share in an expedited manner. The number next to the price is the available volume at this price, also called depth. 

The concepts of spread and mid-price are important in the orderbook. Spread refers to the distance between the best ask price and bid price, and the mid-price is the average price of the best ask price and best bid price. On the one hand, the spread can be explained as a measurement of liquidity. When comparing different markets, the market with more trading volume has a narrower spread. In Chapter \ref{cha:meth}, this is also observed in China commodity markets where we compared with different future instruments in figure (\ref{fig:spreadTurnover}). On the other hand, the spread can also be viewed as the potential reward and risks faced by market markers as introduced in Section \ref{cha:back:ms}. 

\begin{figure}
	\centering
	\subfigure[Origin orderbook]{
		\begin{minipage}[b]{0.4\textwidth}
			\includegraphics[width=1\textwidth]{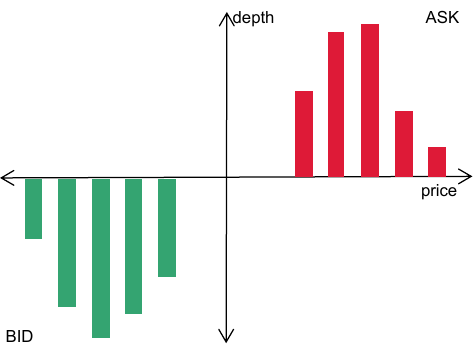} 
		\end{minipage}
		\label{fig:ob1}
	}
    	\subfigure[Orderbook after limit order]{
    		\begin{minipage}[b]{0.4\textwidth}
   		 	\includegraphics[width=1\textwidth]{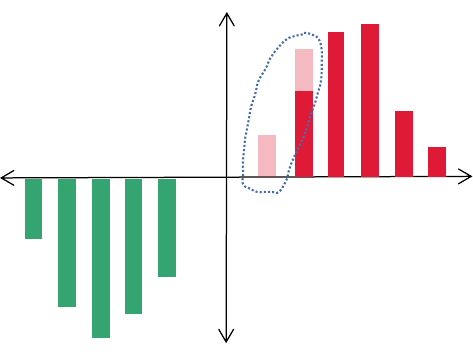}
    		\end{minipage}
		\label{fig:ob2}
    	}
	\\ 
	\subfigure[Orderbook after market order]{
		\begin{minipage}[b]{0.4\textwidth}
			\includegraphics[width=1\textwidth]{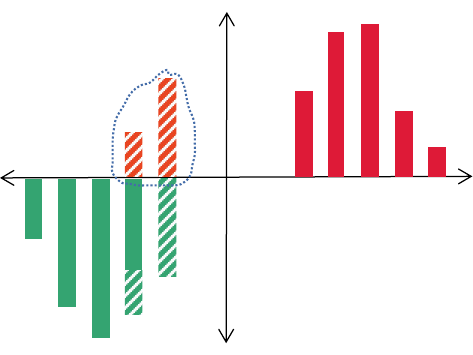} 
		\end{minipage}
		\label{fig:ob3}
	}
    	\subfigure[Orderbook after cancel action]{
    		\begin{minipage}[b]{0.4\textwidth}
		 	\includegraphics[width=1\textwidth]{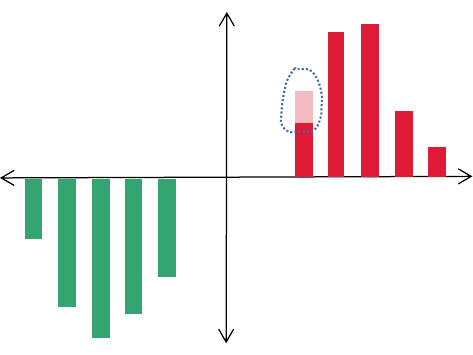}
    		\end{minipage}
		\label{fig:ob4}
    	}
	\caption{Different actions in market}
	\label{fig:obDynamic}
\end{figure}

The group of figures \ref{fig:obDynamic} demonstrates different basic actions applicable to the market. Figure \ref{fig:ob1} depicts the origin orderbook before any actions. If a participator submits limit orders, the orderbook will evolve to the situation in the figure \ref{fig:ob2}. One can see from the figure \ref{fig:ob2} that a new limit order is submitted, and it narrows the spread or increase the depth. In both cases, the liquidity of the market is promoted. If it is better than the current best price, it will push the best price closer to the opponent's side and narrow the spread. Market order are explained in figure \ref{fig:ob3}. If a participator submits a market order or a marketable limit order(a limit order with an executable price), it will be filled based on price-time priority. The orders with the most favorable price will be matched first, and if two orders have the identical price, the order which arrives earliest will be matched first. If the new arrival order is not matched or is just partly matched, the rest of the share will be inserted into the orderbook. From figure \ref{fig:ob3}, a market order/marketable limit order will increase the spread or decrease the depth. Therefore, it will consume the liquidity. The idea of the market order is to guarantee immediate execution regardless of the prices. Figure \ref{fig:ob4} demonstrates the cancel action of a limit order. Participators who submit limit orders are also called 'makers', and participators who submit market order/marketable limit orders are also called 'takers'.

Orderbook reflects the explicit willingness of supplies and demands as well as the liquidity at a given time point. There is also hidden liquidity, including special order types like iceberg orders or just expedited order submission when the favored price is reached. When a trader submits a limit order, he fosters the liquidity of the market. From this end, the potential profit spread is the reward for the market maker's contribution to providing liquidity. 

\subsection{Market Making via Stochastic Control}
\label{cha:pre:MM:type}
The market-making can be divided into stochastic control based\cite{avellaneda2008high, stoikov2009option, korajczyk2019high, cartea2017algorithmic, ho1981optimal, pham2013} and reinforcement learning based\cite{spooner2018, spooner2020robust, zhong2020}. The general framework of RL is elaborated in section \ref{cha:pre:DRL:RL}. Against the market-making background, states can be represented by any predictive features for short trends and other representative variables for the agent's status. Ga{\v{s}}perov exhaustively summarizes RL-based methods from different dimensions in \cite{gavsperov2021reinforcement}. Under a stochastic control framework, the solution heavily relies on assumptions. Here we briefly elaborate on the general framework of stochastic control and explain the idea again market-making background.

Agent's value at time $t$ with state $S_t$ can be represented by value function $H(t,S_t)$ in equation \ref{eq:scVfunction}. 
\begin{equation}
H(t,S_t)=\sup_{\pi\in\mathcal{A}_{0,T}}[\mathbb{E}_{t}(U)]=\mathbb{E}_t(U^{\pi^*}_T)
\label{eq:scVfunction} 
\end{equation}
It equals to expected optimal utility at terminal time $T$ when applied optimal strategy $\pi^*$. $\mathcal{A}$ is the admissible set which refers to all strategies with finite possible loss. Strategy is confined within $\mathcal{A}$, and it rules out strategies like the martingale betting strategy, which may induce unbounded loss. State $S_t$ is $\mathcal{F}$-measurable and $\mathbb{E}_t$ refers to expectation operator $\mathbb{E}_t[\cdot]=\mathbb{E}[\cdot|\mathcal{F}_t]$. More general, when a strategy is given, $H^\pi(t,S_t)=\mathbb{E}_t[U^{\pi}_T]$ refers the value function when strategy $\pi$ is applied.

Then, it can be paraphrased with two time points $H(t_0,S_{t_0})$ and $H(\tau,S_\tau)$ in equation \ref{eq:scVfunctionTwo}. This trick can recursively scale down the origin problem into smaller problems and it is widely-used in dynamic programming.

\begin{equation}
H^\pi(t_0,S_{t_0})=\mathbb{E}_{t_0}[H^\pi(\tau,S_\tau)+\int_{t_0}^\tau F(t,S^\pi_t,\pi_t)dt]
\label{eq:scVfunctionTwo}
\end{equation}

The term of $\int_{t_0}^\tau F(t,S^\pi_t,\pi_t)dt$ refers to the running rewards during the time between $t_0$ and $\tau$. One can find some resemblance this equation and Bellman Equations \ref{eq:policyIter}. When $\tau\rightarrow t^+$, with Ito's lemma, we reach Hamilton-Jacobi-Bellman Equation(HJB):

\begin{equation}\label{eq:HJB} 
\begin{aligned} 
\partial_t H(t,S_t)+\sup_{\pi \in \mathcal{A}}(\mathcal{L}_t^\pi H(t,S_t)+F(t,S_t,\pi))&=0\\
H(T,S_T)&=G(S_T)
\end{aligned}
\end{equation}


$G(S_T)$ is the terminal condition. $\mathcal{L}$ operator is the infinitesimal generator, and its format differs in types of stochastic processes. Since we use a one-dimensional diffusion process and jump process in this section, we listed the infinitesimal generator $\mathcal{L}$ for them.

For diffusion process and counting process, the SDE are listed in formula \ref{eq:SDE:D} and \ref{eq:SDE:C} where the $W_t$ is standard Wiener process. Strictly, $N_t^\pi$ is a counting process with intensity $\lambda^\pi_t$.

\begin{equation}\label{eq:SDE:D} 
dX^\pi_t=\mu(t,x,u)dt+\sigma(t,x,u)dW_t
\end{equation}

\begin{equation}\label{eq:SDE:C} 
    dX^\pi_t=dN_t
\end{equation}

The infinitesimal generator for them are listed in formula \ref{eq:LforD} and \ref{eq:LforC} respectively.

\begin{equation}\label{eq:LforD} 
\begin{aligned} 
\mathcal{L}_t^\pi & =\mu_t^u\partial_x+\frac{1}{2}(\sigma^\pi_t)^2\partial_{xx}\\
 & = \mu(t,x,u)\partial_x+\frac{1}{2}\sigma^2(t,x,u)\partial_{xx}
\end{aligned}
\end{equation}

\begin{equation}\label{eq:LforC} 
\mathcal{L}_t^\pi H(t,n) =\lambda(t,n,\pi)[H(t,n+1)-H(t,n)]
\end{equation}

After reaching HJB equations, which essentially are partial differential equations(PDEs), they can be solved by numerical methods like the finite difference or by analytical solutions with some simplification. Now we discuss stochastic control against the specific background of market-making. We use the method introduced in \cite{avellaneda2008high} as the backbone to model the market making and derive HJB equations. In the order-driven market, the agent thrives on maximizing its wealth by quoting limit with $P_t^a$ and $P_t^b$ at both sides. First, we assume mid-price process $S_t$ as Geometric Brownian motion, accumulation of inventory at ask side and bid side as counting process $N_t^{\pi,a}$ and $N_t^{\pi,b}$ with controlled intensity process $\lambda_t^{\pi,a}$ and $\lambda_t^{\pi,b}$. Then the net inventory process $Q_t$ is $N_t^{\pi,b}-N_t^{\pi,a}$. With the help of the inventory process and quote price process, the cash process can be written as follows:

\begin{equation}\label{eq:MM:cashProcess}
dC_t^\pi=P_t^{\pi,a}dN_t^{\pi,a}-P_t^{\pi,b}dN_t^{\pi,b}
\end{equation}

$\delta_t^{\pi,a}$ and $\delta_t^{\pi,b}$ is the distance from mid-price to ask price and bid price respectively, and they are fully controlled by the agent.

\begin{equation}\label{eq:MM:spread} 
\begin{aligned} 
\delta_t^{\pi,a}&=P_t^{\pi,a}-S_t \\
\delta_t^{\pi,b}&=S_t-P_t^{\pi,b}
\end{aligned} 
\end{equation}

In the study\cite{avellaneda2008high}, the controlled intensity process $\lambda_t^{\pi,a}$ and $\lambda_t^{\pi,b}$ are assumed solely related to $\delta_t^{\pi,a}$ and $\delta_t^{\pi,b}$ respectively. Among those processes, quote price processes are directly controlled by the agent's strategy, and it will affect the inventory process, cash process, and wealth process, whereas the mid-price process is independent of the agent's strategy. Format of value function $H(S_t,t)$ can be parallel as a reward function in RL, and it affects the behavior pattern of agents. The liquidation function represents the wealth if the agent decides to quit the market and liquidates his inventory.

\begin{equation}\label{eq:MM:liquidation} 
    L(c,q,s)=c+q\times s
\end{equation}

The agent's value function is in this format:
\begin{equation}\label{eq:MM:valueFunction} 
H^\pi(C_t,Q_t,S_t,t)=\mathbb{E}_t[U(L(C^\pi_T,Q^\pi_T,S_T))]=\mathbb{E}_t[U(C^\pi_T+Q^\pi_T\times S_T)]
\end{equation}

The SPDE for mixed diffusion and jump process is:
\begin{equation}\label{eq:MM:dh}
    dH_t^\pi=\mu_t^\pi dt+ \sigma_t^\pi dW_t+\gamma_t^{\pi,a}dN_t^{\pi,a}+\gamma_t^{\pi,b}dN_t^{\pi,b} 
    \end{equation}

And the corresponding HJB equations for is:

\begin{equation}\label{eq:MM:hjb}
\begin{aligned} 
    H_t^\pi+\frac{1}{2}\sigma^\pi_t H_{ss}& +\max_{\delta^{b}}\lambda^b(\delta^b)[H(s,c-s+\delta^b,q+1,t)-H(s,c,q,t)]\\
    & +\max_{\delta^{a}}\lambda^a(\delta^a)[H(s,c+s+\delta^b,q-1,t)-H(s,c,q,t)]=0\\
    &H(s,c,q,T)=U[c+s\times q]
\end{aligned} 
\end{equation}

By replacing utility function $U(\cdot)$ and intensity $\lambda(\cdot)$ with specific form. This control problem is converted into a PDEs. In the original study, Avellaneda gives analytical solutions with first-order approximation. Numerical solution methods like finite difference are also widely-used. 

The stochastic control method has its limitations. First, Some assumptions are rigid. For example, the intensity of the inventory counts process does not solely depend on the self side but also depends on the opponent side. Besides, how incorporating other essential variables, including commission fees, rebates, and cancel limitations, is also hard under this framework.


\chapter{Methodology}
\label{cha:meth}
To our best knowledge, our study is the first to apply deep reinforcement in the China Commodity Market. In this study, we develop the ATS for the China Commodity Market. Besides, we develop the market making via reinforcement learning and compare it with two baseline models. Last but not least, we analyze the behavior pattern under different market environments. The chapter is organized as follows: the structure of our ATS is explained in Section \ref{cha:meth:ats}. Also, in Section \ref{cha:meth:ats}, we conduct latency analysis to test the inner latency and external latency of our ATS. Section \ref{cha:meth:data} demonstrates the data used in this thesis, and Section \ref{cha:meth:env} illustrates each components of RL and environment. Section \ref{cha:meth:expDesign} explains the experiment settings and the evaluation metrics.

\section{Developing ATS for China Commodity Market}
\label{cha:meth:ats}
\begin{figure}
    \centering   
    \includegraphics[]{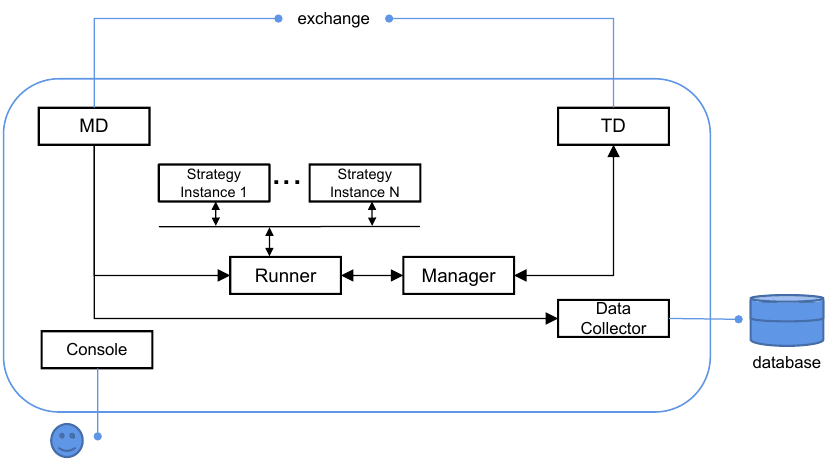}
    \caption{Component figure for Xtrader ATS}  
    \label{fig:component}  
\end{figure}

For a high-frequency trading strategy, the automatic trading system(ATS) is a key IT infrastructure and is at least as important as the strategy itself. ATS response for multi functionalities, including the communication between strategic algorithm agents and exchanges, the management of different trading strategies, and low-level technical details. In a quantitative hedge fund, a well-functioned team contains teammates with different backgrounds, and not all of them are experts in low-level techniques like communication techniques and high-efficiency computing. Besides, the interface provided by the exchanges could be complicated and troublesome for strategy developers to use. In this study, to test the RL agent and analysis its performance from an empirical perspective, a customized trading system Xtrader has been developed. Xtrader is developed with mixed Python and C++ languages and contains multi-components that are implemented in an asynchronous manner. Also, it can function as a platform that provides a convenient interface for strategy developers and can be deployed different strategies. Multi-strategy can help to diversify capital allocation and reduce the risks. The communication between different services is through shared memory and enjoy a superior advantage in speed over other communication mechanisms like storage I/O or web service. The communication protocol resembles to message queue. For each queue, the only controller can write it while the multi subscribers can read it. The writing and reading are asynchronous; therefore, the writer will not need to wait for any return. For now, Xtrader is successfully deployed and receives 8 million records every trading day, which takes up 1-gigabyte storage.
Figure \ref{fig:component} demonstrates the architecture of our ATS. It contains five components of service: Console, MD/TD, Runner, Manager, and Collector.

The console is responsible for controlling the activation of other modules and is the place human administrator can control the start and end of this ATS. MD/TD services will respond to the low-level communication with exchanges. MD receives the real-time market data and writes it into shared memory. TD is responsible for sending actions to exchanges and handling the returns from the exchanges. The runner is the execution engine for registered inner strategies. It is designed to support multi-strategies. What is more, it provides a convenient high-level interface for developing the strategies. With the provided interface, the strategy instance can conveniently give basic orders and advanced orders. Basic order contains limit order and cancels action, which is introduced in \ref{cha:pre:MM:ob}. Also, some predefined advanced orders like iceberg order can save the trouble for the strategy when the operation is complex. For example, a typical composite order is the stop order, in which open price, stop profit and stop loss need to be provided.

The manager module is responsible for translating inner requests to the specific format defined by China commodity futures exchanges. It also checks the legitimacy of the order. In this manner, functions related to exchanges are decoupled, and Xtrader can be easily extended to support other exchanges like Binance for Crypto and Multi Commodity Exchange of India Limited(MCX) for the commodity in India, where the trading rule sets will be different. Only minimum modifications need to make to the Manager module when migrating to other markets. The Collector collects the data received by MD in shared memory and writes it into the database. Other information like whether it is a holiday or the liquidity ranking are also updated into the database through Collector. The current database only contains rudimentary data types. In the future, information like Candlestick chart or some relevant web information like warehouse receipts should also be collected and maintained. MD/TD are implemented with C++ Because they need to handle relatively large data throughout. Temporally, other modules are implemented by Python.


\begin{table}
	\centering
	\begin{tabular}{{c|c|c}} \toprule
	   & Inner Latency & External Latency \\ \hline
	Mean($ms$)  & 0.40 & 35.22  \\ \hline
	Min($ms$) & 0.25 &  31.36 \\ \hline
	Max($ms$) & 1.27&  332.06 \\ \hline
	Std($ms$) & 0.35 & 16.36 \\  \hline
	Sample Size & 2000 & 2000  \\ \bottomrule
	\end{tabular}
	\caption{Inner and external latency analysis}
	\label{tab:lantency}
\end{table}

\begin{figure}
	\centering
	\subfigure[Inner latency plot]{
		\begin{minipage}[b]{0.45\textwidth}
			\includegraphics[width=1\textwidth]{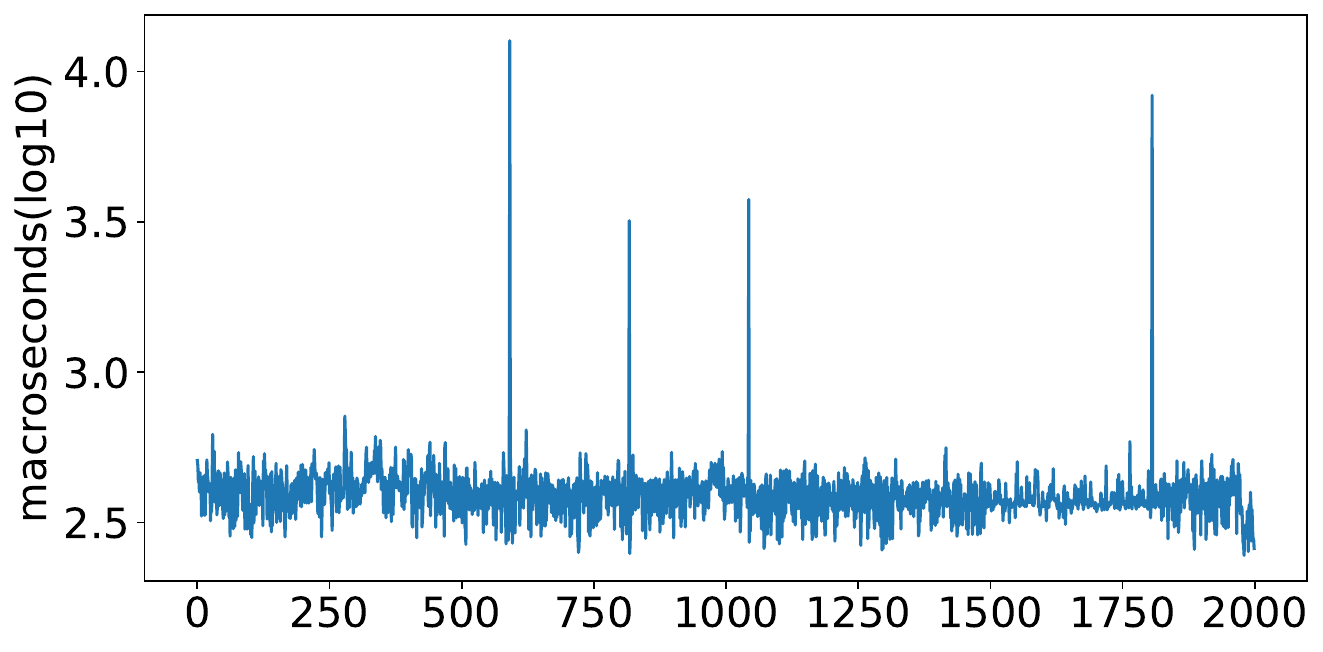} 
		\end{minipage}
		\label{fig:innerLatency}
	}
    	\subfigure[External latency plot]{
    		\begin{minipage}[b]{0.45\textwidth}
   		 	\includegraphics[width=1\textwidth]{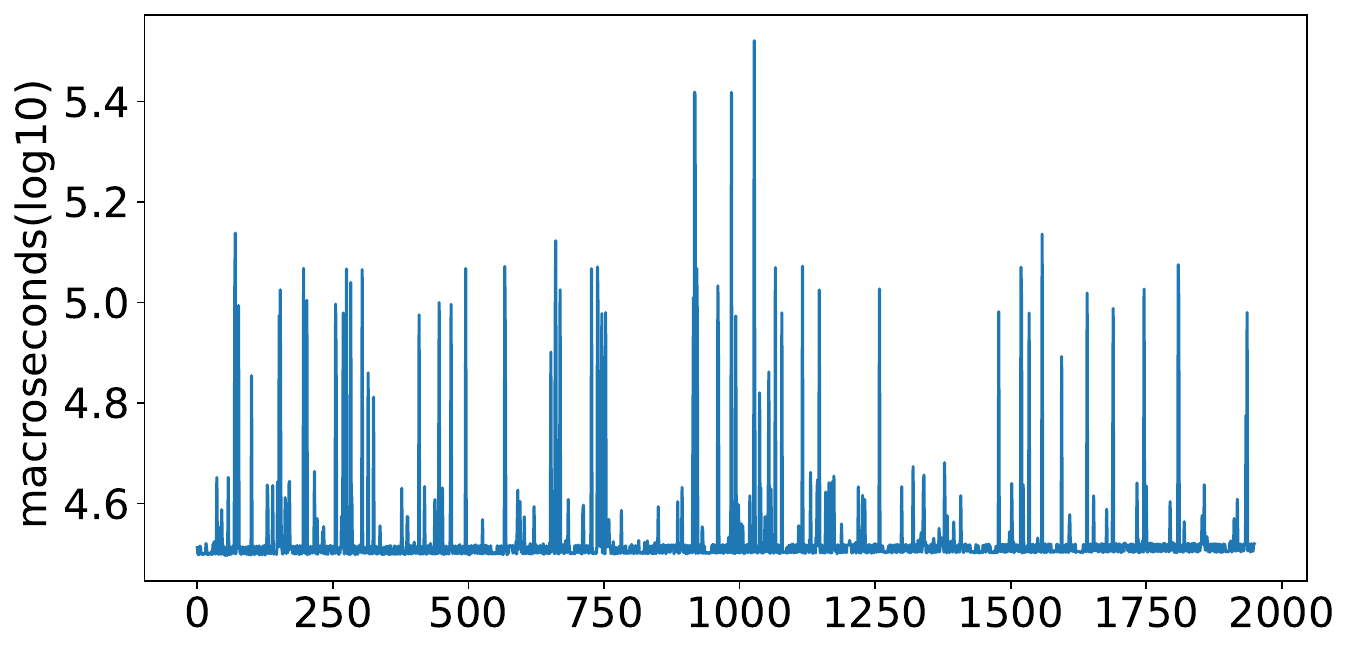}
    		\end{minipage}
		\label{fig:externalLatency}
    	}
\end{figure}

A performance test on Xtrader's latency has been conducted. We built a simple speed test strategy. The strategy will submit limited orders at the best price and cancel them immediately, which puts a very high demand on the efficiency of the ATS. For every 500ms, the agent will cancel the order and resubmit the limit order at the best price. This strategy has been run to accumulate $2000$ submission. Table \ref{tab:lantency} summarizes the inner and external latency. The average inner latency is 400 us, while the external latency is $35$ ms. This result is still far slower than the nano-level trading system used in the first-tier quantitative hedge fund. However, when compared to the external latency(web latency), which is $35$ ms, this is temporarily negligible. Therefore, the current bottleneck is the web service which can be easier solved if it is deployed on co-location. In the future, we can improve the performance of Xtrader by re-implementing it in C++ or using FPGA.

\begin{table}
\centering
\begin{tabular}{@{}llr@{}} \toprule
	Fields    & Description \\ \midrule
	update time & the update timestamp \\
	ask price 1-5  & the best price at ask side     \\
	bid price 1-5		& the best price at bid side         \\
	ask volume 1-5 & depth for ask prices     \\
	bid volume 1-5 & depth for bid prices     \\ 
	last price       & the latest executed price      \\
	volume       & the number of traded volume      \\
	open interest       &   total number of outstanding derivative contracts    \\
	turnover& the amount of traded volume in CNY  \\ \bottomrule
\end{tabular}
\caption{Field\& description for raw data}
\label{tab:dataRecord}
\end{table}

\section{Data Collection and Derivative Features}
\label{cha:meth:data}

The trading rule differs in exchanges, and so does the data disclosure. For trading rules and regulation articles in the China Commodity market, they are articulated on the official website\cite{shanghai2021exchange}. In the China Commodity futures market, the data is provided in an aggregated manner on a periodic basis. For every 500 milliseconds, the exchanges will push the data to public channels. Tables \ref{tab:dataRecord} represent the fields of released data. Update time is the timestamp generated by exchanges which can also be viewed as the identity of a record. Ask prices $1$ to $5$ and bid prices $1$ to $5$ are the best price at the ask side and bid side. Volume refers to the trading volume in the last 500 milliseconds.

\begin{figure}
	\centering
	\subfigure[Prices of rebar's different instruments]{
		\begin{minipage}[b]{0.45\textwidth}
			\includegraphics[width=1\textwidth]{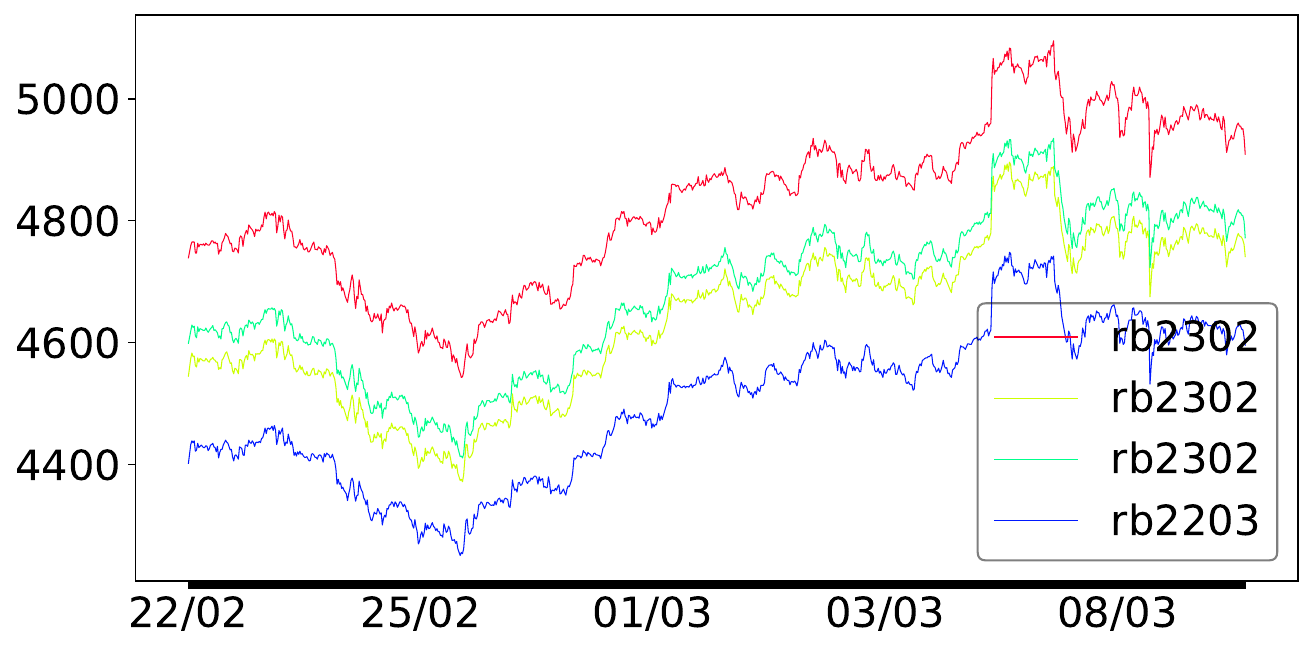} 
		\end{minipage}
		\label{fig:coninter}  
	}
    	\subfigure[Relationship between spread and turnover(in log(10))]{
    		\begin{minipage}[b]{0.49\textwidth}
   		 	\includegraphics[width=1\textwidth]{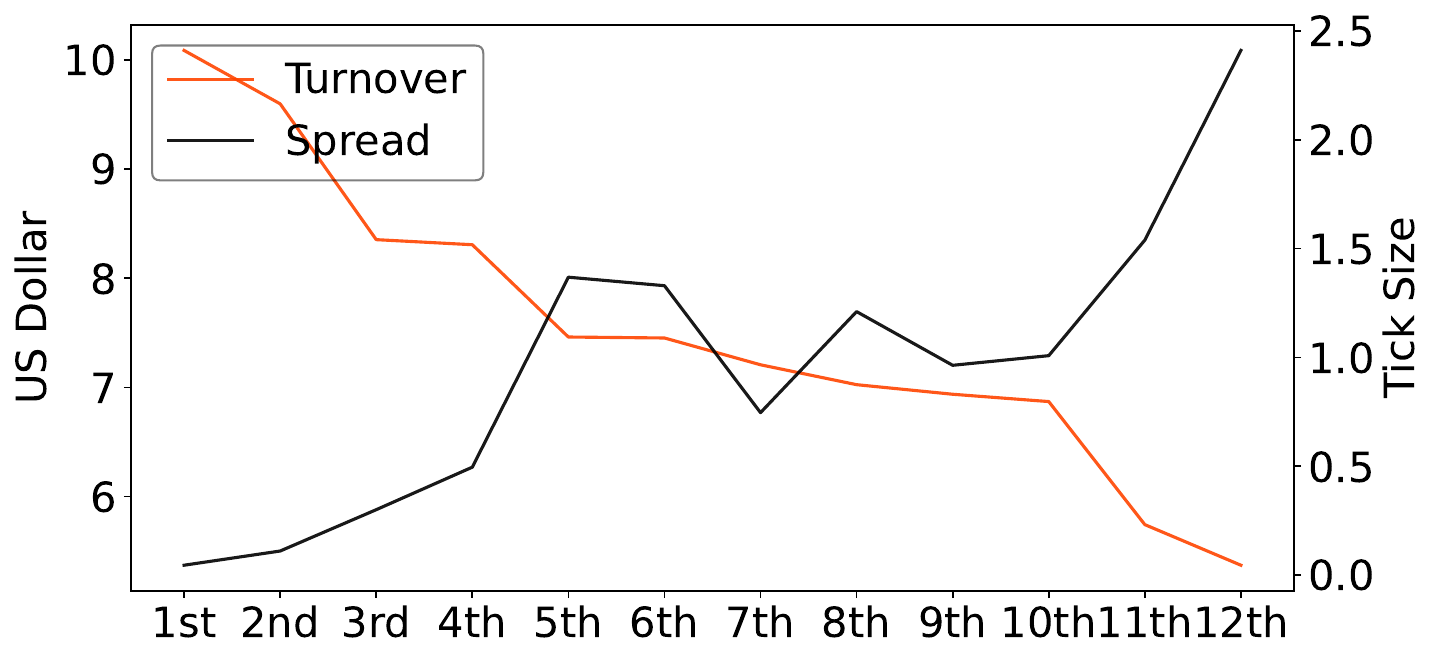}
    		\end{minipage}
		\label{fig:spreadTurnover}
    	}
\end{figure}

For each commodity product, there are multi instruments with different maturities. Reinforcing bar(also known as 'rebar'), for example, is an active trading commodity in China and has 12 derivatives with maturities for the next 12 months. From Feb 2022 to Mar 2022, the average daily trading turnover for rebar is 13 billion measured in US dollars(\$). The naming rule for the instrument is $product+maturity$. For example, The 'rb2207' is the rebar commodity future derivative which will be mature in July 2022. The price of those instruments with the same underlying assets demonstrates highly co-trend as shown in \ref{fig:coninter}. The price difference between two instruments can be expressed by $Y_t^i-Y_t^j=\beta_t+\varepsilon_t$. $Y_i$ and $Y_j$ refer to two instruments with different maturities. The beta term is the system difference term which is relatively stable and may reflect the public expectation of future supply-demand relationship or cost/profit of carrying spots. The $\varepsilon_t$ is the error that can be attributed to random factor or temporal dispersion. 

Instruments with the same underlying commodity but different maturity have different liquidity and spread. Figure \ref{fig:spreadTurnover} demonstrates the relationship between the daily turnover in US dollars and the spread. One can see spread negatively correlates with turnover. As discussed in Section \ref{cha:pre:MM}, the reward of the market maker can be viewed as compensation for the liquidity he provides to the market. Therefore, a relatively less active market might be more profitable to conduct market-making. Also, the commission fee should also be taken into account. For rebar, the current commission fee for the public is 2$\%\%$ of the amount of the total contract value, which amounts to around one tick price. Therefore, rebar instruments whose spread is smaller than two are not profitable without predicting price trends.
%

\begin{figure}
	\centering
	\label{fig:coef}
	\subfigure[Correlation of mid-price]{
		\begin{minipage}[b]{0.47\textwidth}
			\includegraphics[width=1\textwidth]{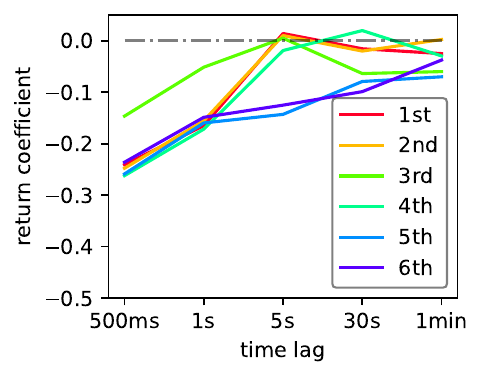} 
		\end{minipage}
		\label{fig:coefTimelag}
	}
    	\subfigure[Correlation of last price]{
    		\begin{minipage}[b]{0.47\textwidth}
   		 	\includegraphics[width=1\textwidth]{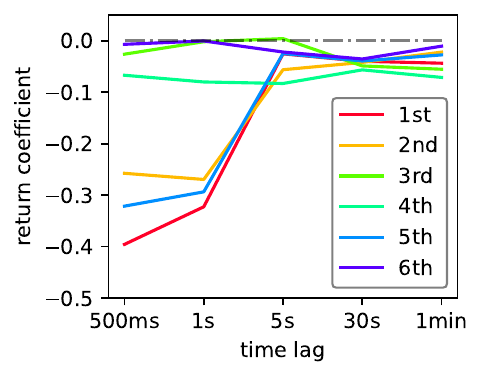}
    		\end{minipage}
		\label{fig:coefTimelagLastPrice}
    	}
\end{figure}

We also investigate the efficiency of the market in the high-frequency world. Figure\ref{fig:coef} demonstrate the first-order auto-correlation coefficient in different timescale. When the sampling frequency increase, the negative auto-correlation coefficient of the return series becomes increasingly significant. Result reputes the idea that the process of last price or mid-price is perfectly martingale, at least not the truth in the high-frequency setting. The returns series demonstrate the feature of mean-reverting, and part of the price is predictable. Therefore, if the market maker only takes inventory control as his sole purpose, he will be largely exposed to adverse selection risks and suffers from loss. However, adverse selection risk is ignored by the stochastic control-based method introduced in \ref{cha:pre:MM:type}. How to incorporate predictive signals into market makers is viable in the success of market makers. 



\section{Environment Setting and State Engineering}
\label{cha:meth:env}
\begin{figure}
    \centering   
    \includegraphics{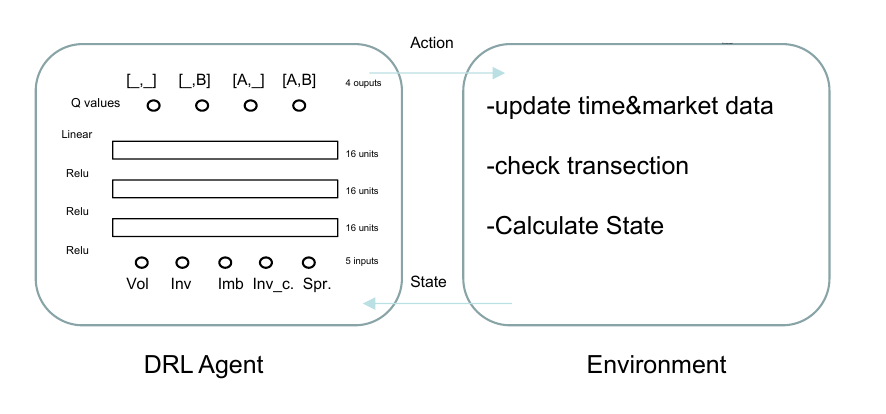}
    \caption{Overview of the interaction between the DRL agent and the environment}  
    \label{fig:agentEnv}  
\end{figure}

Figure \ref{fig:agentEnv} illustrates the interaction between the DRL agent and environment. For deep reinforcement learning, the five most important components are action space, state space, reward function, function approximator, and learning algorithm. Function approximator and learning algorithm have been introduced in Section \ref{cha:pre:DRL}. Also, for market making, since the execution of a limit order is a probabilistic event, the assumption of execution is also important. The followings are the explanation for the other three components and execution mechanisms.

\subsection{Action Space}
\label{cha:meth:env:as}
Action space $\mathcal{A}$ is the collection of all possible actions that an agent can take. On each side, the agent can only pose limit order or no action. Limit action will pose limit order on the best price on the self side, and it waits for the arrival of market order from the opponent side. $None$ means that there would be no order at this timestep. Actions will be updated for every time step and will expire at the end of this time step. Volume for limit orders is default unit one. Therefore, the action space $\mathcal{A}$ contains 4 actions: $(N,N)$, $(N,B)$, $(A,N)$, $(A,B)$.

\subsection{Execution Mechanism}
\label{cha:meth:env:execution}

\begin{figure}
    \centering   
    \includegraphics{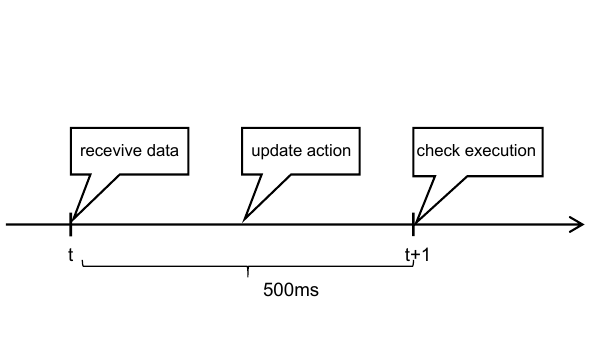}
    \caption{Timing mechanism}  
    \label{fig:timing}  
\end{figure}

Since limit order cannot guarantee execution, another important aspect is the assumption about timing and execution. Figure \ref{fig:timing} illustrates the timing. It is assumed that after receiving data at $t$, an agent should submit the orders before the next timestamp, and whether it will be executed will be checked at $t+1$ when the new market data is updated. A limit order is executed if the last price of next step $t+1$ has crossed limited price at $t$, that is $last\_price_{t+1}>ask\_price_t$ for ask limit order and  $last\_price_{t+1}<bid\_price_t$ for bid limit order. And both the cash process $C_t$ and inventory value process $Inv_t$ will change, and so does the wealth process. The cash process, inventory value process, and wealth process are given below:

\begin{equation}
\begin{aligned}
	& \Delta C_t =\frac{1}{2}(match\_LO^a_{t-1}+match\_LO^b_{t-1}) \\
	& \Delta Inv = -match\_LO^a_{t-1}+match\_LO^b_{t-1} \\
	& \Delta Inv\_value =Inv_t*(mid\_price_t-mid\_price_{t-1}) \\
	& \Delta Wealth_t =\Delta C_t+\Delta Inv\_value_t
\end{aligned}
\end{equation}

Then cash process will increase cash by the amount of $\frac{1}{2} Spread_t$ when a limit order is executed. The change of inventory depends on the quote side. Execution on the ask side will decrease the inventory by a unit, while an order executed on the bid side will increase the inventory by a unit.




\subsection{State Space}
\label{cha:meth:env:ss}
State-space $S$ is the collection that aggregates all representative and predictive features for the market. Table \ref{tab:stateVarible} lists states used in this study. It contains inventory, volatility, inventory value change, spread, and orderbook imbalance. Inventory of the agent and spread is directly relevant to marking making since it determines the inventory risk and the cost of limit order. Volatility is also included because they are also relevant to market-making from our discussion in Section \ref{cha:back:ms}. Besides inventory, we also include orderbook imbalance which is a predictive indicator of short-term price changes. Finally, the change in inventory value is also included since it may reflect the adverse selection risk it suffers from its informed rivals.

\begin{table}
	\centering
	\begin{tabular}{@{}llr@{}} \toprule
	State Variable   & Description & References \\ \midrule
	Orderbook Imbalance  & $$$
	(\sum_{i=1}^5 V_i^b-\sum_{i=1}^5 V_i^a)/(\sum_{i=1}^5 V_i^b+\sum_{i=1}^5 V_i^a)
	$$$ & \cite{zhong2020}    \\

	Volatility		& $std[MP_{t-l}:MP_t]$   & \cite{haider2019effect}      \\
	Inventory & size of inventory  & \cite{zhong2020,spooner2018}  \\
	Spread & size of spread    &  \cite{zhong2020,mani2019applications} \\ 
	Delta of Inventory Value  & $ I_t-I_{t-50}$  &  \cite{zhong2020}     \\ 
	
	\bottomrule
	\end{tabular}
	\caption{State variables used in this study}
	\label{tab:stateVarible}
\end{table}

\subsection{Reward Function}
\label{cha:meth:env:reward}

Reward functions are used to guide the RL agent in training. It is natural to use the return in capital gain $\Delta Wealth_t$ as the reward, which is wide-used in formal studies such as \cite{spooner2018, zhong2020}. Asymmetric reward functions are introduced in \cite{spooner2018} to encourage agents to focus on earning the spread instead of catching the trend. The agent will get punishment for loss in inventory value change but will not get the reward from the profit in inventory value change. The inventory punishment term can be applied to encourage agents to keep the inventory around zero. The formula for them is given below:

\begin{equation}\label{eq:reward}
\begin{aligned} 
sym\_reward_t & =\Delta Wealth_t =\Delta C_t + \Delta Inv\_value_t  \\
rewardWithInvPunish_t & =\Delta C_t + \Delta Inv\_value_t - punish\_coef*abs(Inv_t) 
\end{aligned}
\end{equation}

\section{Experiment Design and Evaluation Metrics}
\label{cha:meth:expDesign}
So far, the mechanism of market-making and reinforcement learning algorithm has been introduced above. The parameter used in this experiment has been listed in table \ref{tab:parameters} in Appendix \ref{app:A} for reproduction purpose.
To verify the effectiveness of Reinforcement Learning in market making, we compare the performance of the Deep Reinforcement Learning Agent(DRLA) with two baseline models. The first base model is a Fixed Agent(FA), which will constantly quote at the best price. Since the inventory control has been proved an important factor in making, a max constraint on the inventory size has been applied to the fixed model to avoid the inventory increases unlimitedly. This agent is referred as FAwC. Specifically, when the inventory goes to $\pm5$, the action would be modified to avoid continuously accumulating inventory. For example, the action $(A,B)$ would be modified to $(N,B)$ when the inventory is $-5$. The experiment result shows this simple constraint can largely improve the baseline models, which again supports the idea that maintaining the inventory around zero is profitable. The commission fee is also under consideration.

Agents are evaluated with annualized Sharp, trading times, average profit, and imbalance of inventory control. Sharp has widely used metrics in the evaluation of asset returns, and we define imbalance of inventory in this study to reflect the overall exposure direction. Its formula is $\ln(\frac{\int (Inv_t)^+ dt}{\int (Inv_t)^-dt})$ which is the logarithm of the ratio of upper area to bottom area in inventory process curve. When it equals 0, it means the agent has equal exposure to the long side and short side in this trading session. When an imbalance of inventory is larger than zero, the agent is exposed to the risk of price decrease. 

\begin{figure}
    \centering   
    \includegraphics[width=0.95\textwidth]{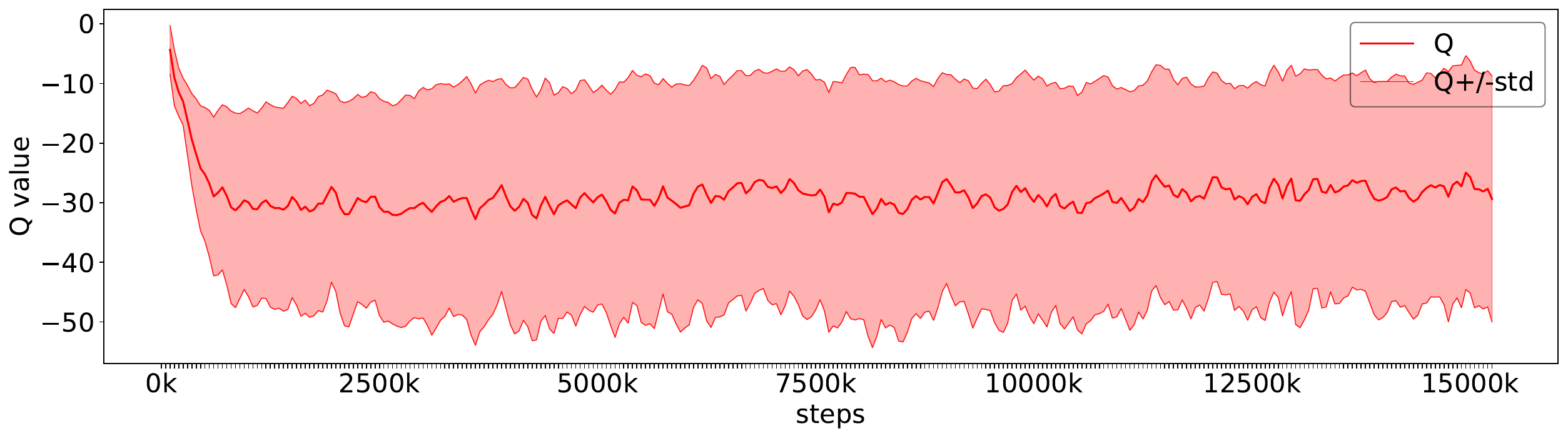}
    \caption{Prices of rebar's different instruments}  
    \label{fig:converge}  
\end{figure}

We use the high-frequency quote data from instruments 'rb2301' from 20220212 to 20220228 as datasets. The whole dataset is divided into a training set, validation set, and set that counts for 5, 2, and 5 trading days. The model was firstly trained over the training set for 30 epochs. For every 5000 steps, the model is evaluated on the validation set, and the best model found on the validation set will be used in the test set. We use annualized Sharp ratio with a constraint on the minimum trading number as the evaluation metrics for model selection on the validation set. Figure \ref{fig:converge} demonstrates the convergence process of the agent's Q value. After a certain training step, the Q value process becomes stationary(strictly at least weak stationary) since the mean function and variance function become stationary. We put minimum constraints on the training times in model selection on the validation set. There are two reasons to support this: 1) the average profit for market making is thin, and only when the trading frequency is high enough the strategy would be practically useful. 2) fewer training numbers would make the evaluation unreliable from a statistical perspective. Therefore, on the validation set, only the model with a daily trading number larger than 50 would be kept.

\chapter{Experiment and Analysis}
\label{cha:exp}
In section \ref{cha:exp:compare}, we verify the effectiveness of RL methods in the market making by comparing it with a baseline model. The partial dependence of actions on each state variable has been analyzed in section \ref{cha:exp:PDL} to analyze the behavior patterns of RL agents. It reveals some interesting patterns and some of which align with market microstructure theory. In section \ref{cha:exp:MM:adv}, we end by summarizing the features of being a successful market maker.



\section{Comparison Experiment}
\label{cha:exp:compare}

The results for comparison are summarized in table \ref{tab:performance}. The comparison experiment suggests the RL is more effective than the two fixed models in both sharp ratio and inventory control. DRLA achieves the highest Sharp and Imb ratio closest to zero. Also, DRLA agent is more prudent in trading and has fewer trading times. Although, because of the black box nature of Deep Learning, we do not know the reason agent's behavior, it is fair to ask if the agent has its judgment of the current market situation? Indeed, we find some interesting behavior patterns when conducting partial dependency analysis in session \ref{cha:exp:PDL}. There are some interesting findings when comparing two baseline models. First, both fixed agents have an Imb ratio significantly smaller than zero, which proves they are exposed to adverse selection risk of negative direction in this test period. Since they quote symmetrically and receive orders passively, the negative imbalance values simply reflect the fact that the market(aggregation of other participators) is initially selling their inventory to fixed agents in this test period. FA has zero intelligence in inventory management and has most deviated Imb value and worse results. However, by simply applying inventory constraint in FAwC, the Imb value can be improved close to zero, and better Sharp can be achieved. It enlightens that putting inventory max constraint is a practical tip in practice to limit both adverse/inventory risks. 

\begin{figure}
	\centering

	\subfigure[DRLA's performance on testset]{
		\begin{minipage}[b]{0.95\textwidth}
			\includegraphics[width=1\textwidth]{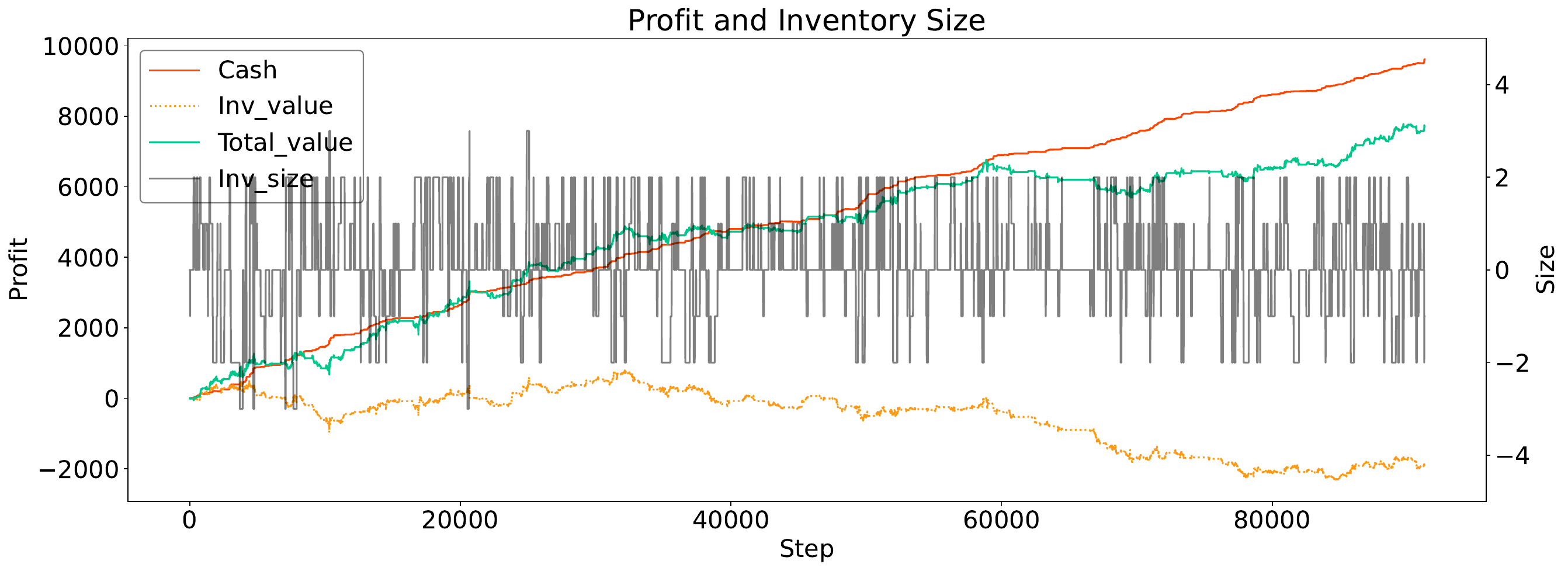} 
		\end{minipage}
		\label{fig:perf:dla:all}
	} 
	\\
	\subfigure[FA's performance on testset]{
		\begin{minipage}[b]{0.95\textwidth}
		\includegraphics[width=1\textwidth]{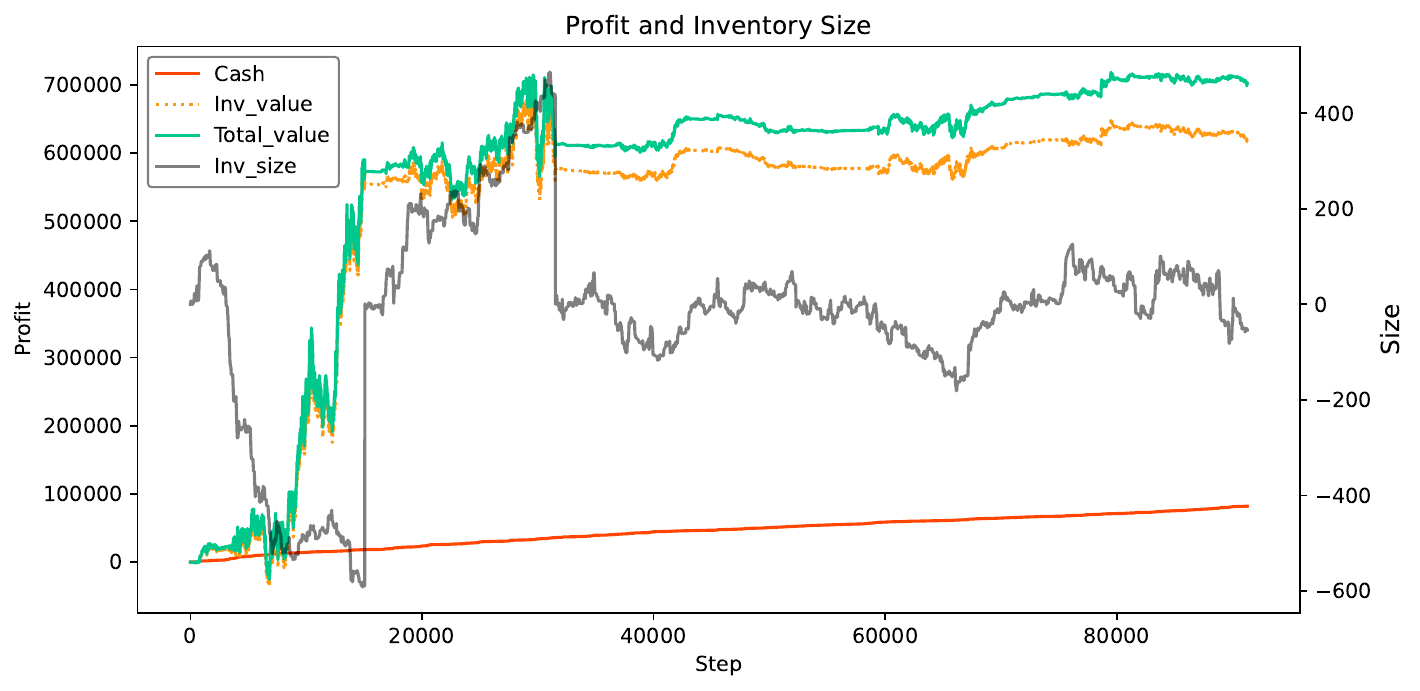}
		\end{minipage}
	\label{fig:perf:fa:all}
	}
	\\
	\subfigure[FAwC's performance on testset]{
		\begin{minipage}[b]{0.95\textwidth}
		\includegraphics[width=1\textwidth]{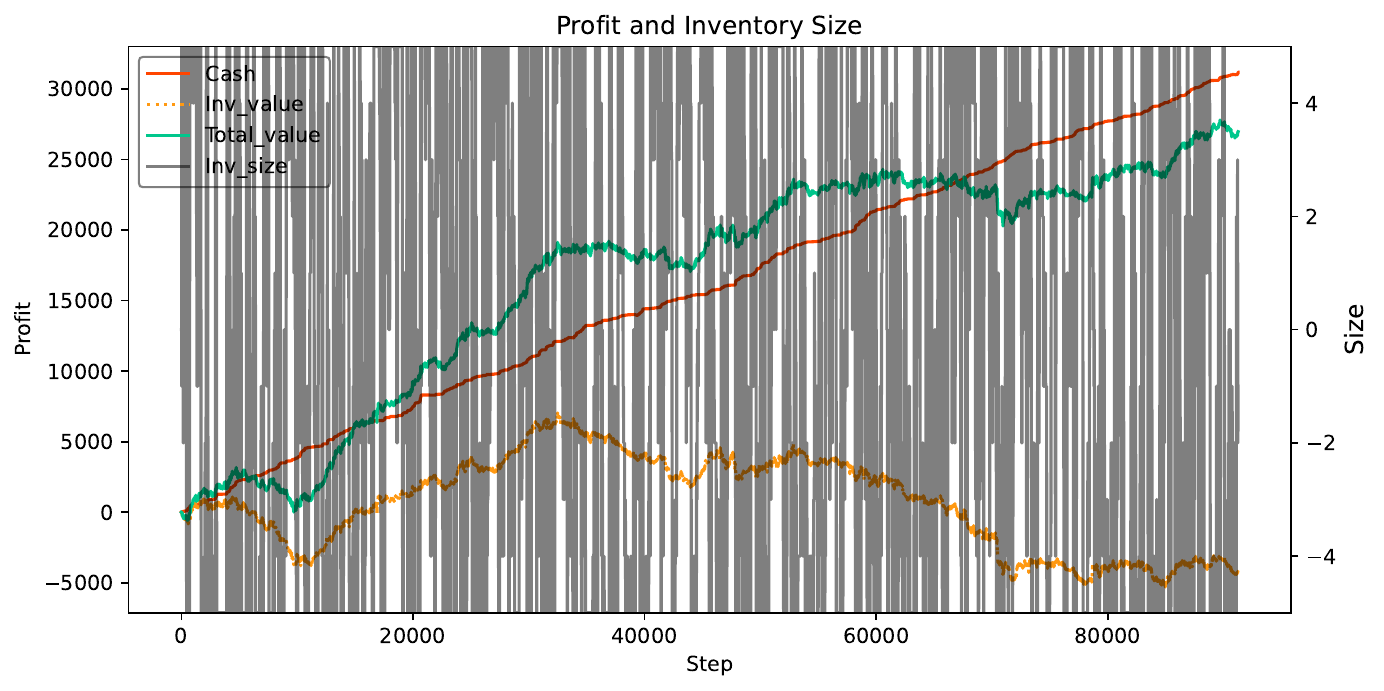}
		\end{minipage}
	}
	\label{fig:perf:fawc:all}

	\caption{Comparison experiment on the testset from 22nd Feb to 28th Feb}
	\label{fig:comparsionAll}
\end{figure}

\begin{figure}

	\subfigure[DRLA's detail actions on the 28th Feb]{
		\begin{minipage}[b]{0.95\textwidth}
			\includegraphics[width=1\textwidth]{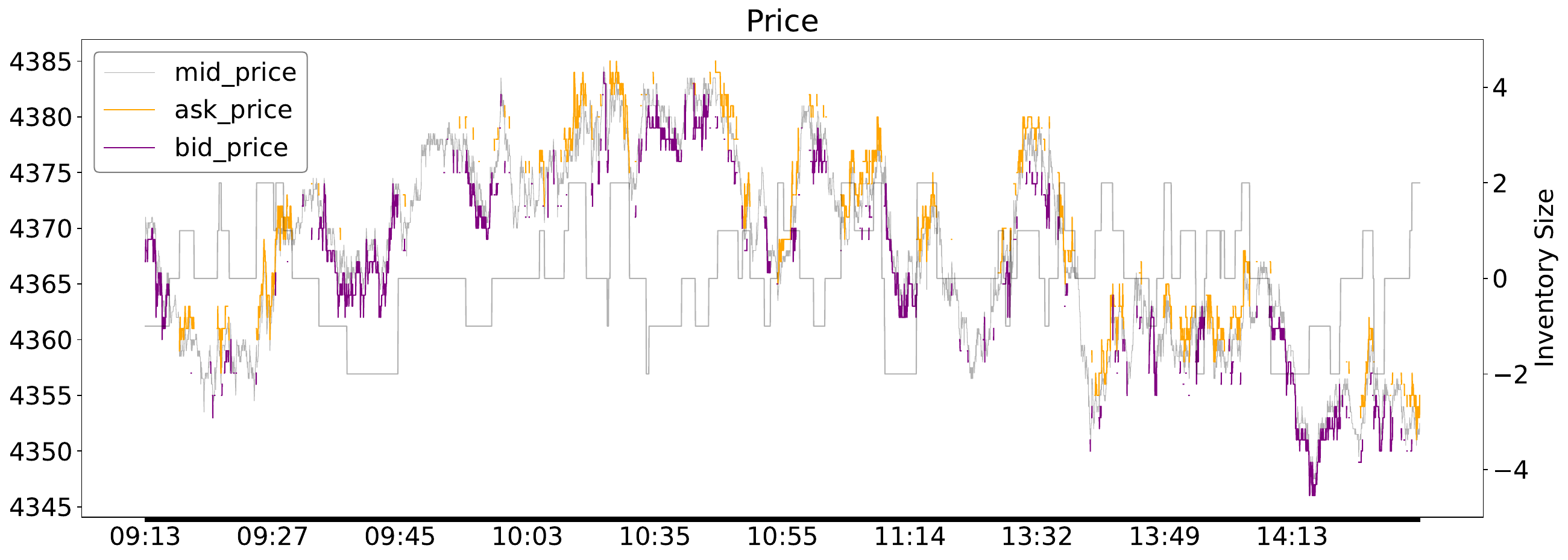} 
		\end{minipage}
		\label{fig:perf:dla:detail}
	}
	\\
	\subfigure[FA's detail actions on the 28th Feb]{
		\begin{minipage}[b]{0.95\textwidth}
			\includegraphics[width=1\textwidth]{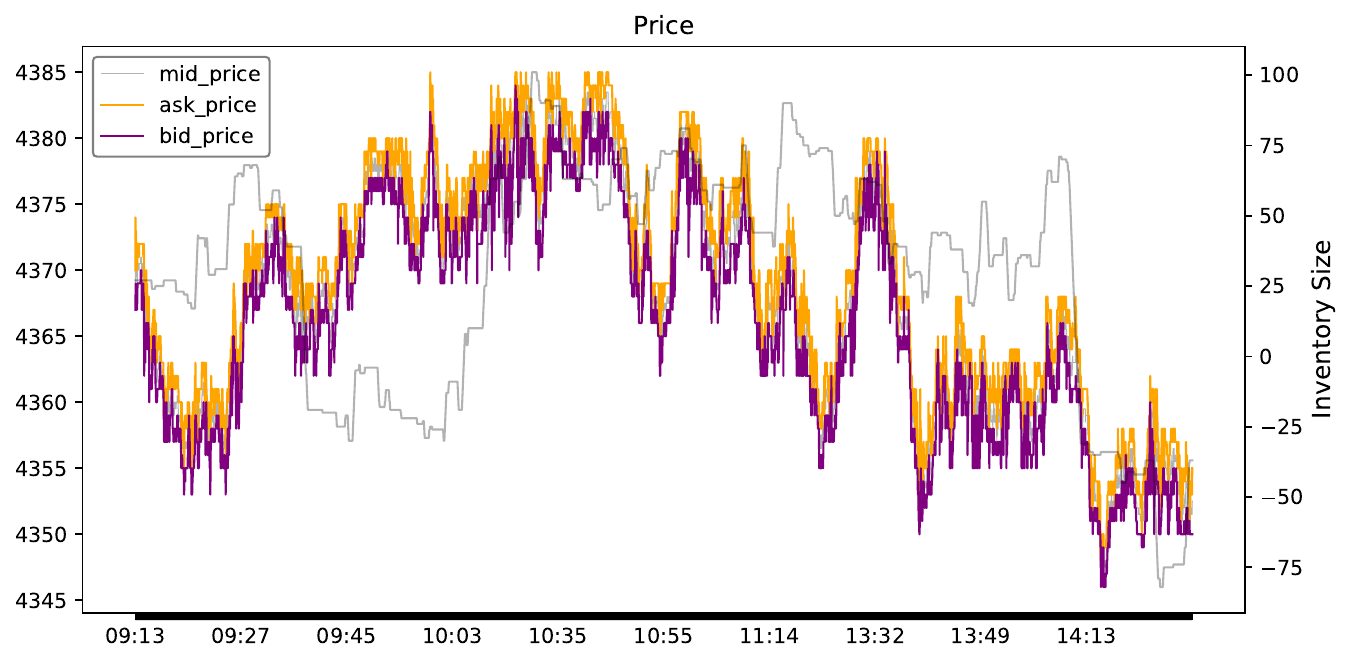} 
		\end{minipage}
		\label{fig:perf:fa:detail}
	}
	\\ 
	\subfigure[FAwC's detail actions on the 28th Feb]{
		\begin{minipage}[b]{0.95\textwidth}
			\includegraphics[width=1\textwidth]{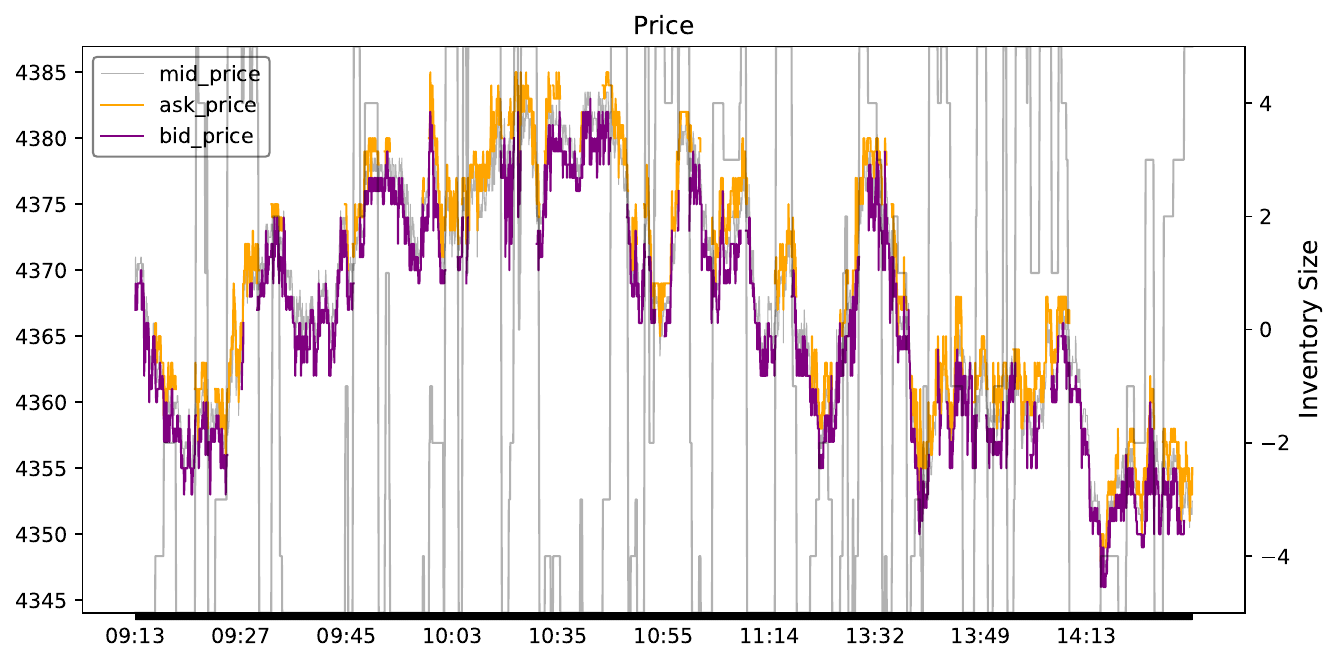} 
		\end{minipage}
		\label{fig:perf:fawc:detail}
	}

	\caption{Detailed performance of agents on an example period(Morning on 28th Feb)}
	\label{fig:comparsionDetail}
\end{figure}

The Profit and Loss(PnL) curve for DRLA agent on testset are given in Figure \ref{fig:perf:dla:all}. As introduced above, the total wealth can be divided into the cash process and inventory value process. Also, They are separated and plotted, and so does the inventory process. From this figure, the DRLA can control the inventory within $\pm 3$ without any explicit constraint on inventory which reflects the effectiveness of this method in inventory control. To demonstrate the detailed control process, we arbitrarily select a half-day episode on the trading session of 28th February(Figure \ref{fig:perf:dla:detail}). Agent quotes strategically and successfully maintains the inventory process around 0. DRLA agent's quote behavior adjusts to the inventory level strategically; when the inventory is accumulated on the long side, the agent tends to quote on the ask side to liquidate the position.


\begin{table}
	\centering
	\begin{tabular}{{c|c|c|c}} \toprule
	   & FA & FAwC & \textbf{DRLA} \\ \hline
	Sharp & 0.01 & 0.82    & \textbf{2.08}  \\ \hline
	Imbalance Ratio  & -0.37 & -0.10   & \textbf{-0.01}  \\  \hline
	Avg. Profit(CNY) & 61.86 &  5.66   & \textbf{2.09}  \\ \hline
	Trading Times & 11348 &  4763   & \textbf{1467} \\ \bottomrule

	\end{tabular}
	\caption{Performance for types of agent}
	\label{tab:performance}
\end{table}

\section{Performance Pattern Analysis}
\label{cha:exp:PDL}
To further investigate the behavior patterns for DRLA and explain(at least partially) decisions of agent's action, the Partial Dependence Plot(PDP) has been made to demonstrate the relationship between individual states and action selection. The mechanism of the partial dependence function is given in the formula below:
\begin{equation}\label{eq:pdp}
\begin{aligned}
h_{x}(a) & =P(A=a|X=x)=\frac{\sum_{Y}P(a,x,Y)}{\sum_{Y,A}P(A,x,Y)} \\
\hat{h}_x(a) & =\frac{\#N(a,x,\cdot)}{\#N(\cdot,x,\cdot)} 
\end{aligned}
\end{equation}

\begin{figure}
	\centering
	\subfigure[Inventory Size]{
		\begin{minipage}[b]{0.47\textwidth}
			\includegraphics[width=1\textwidth]{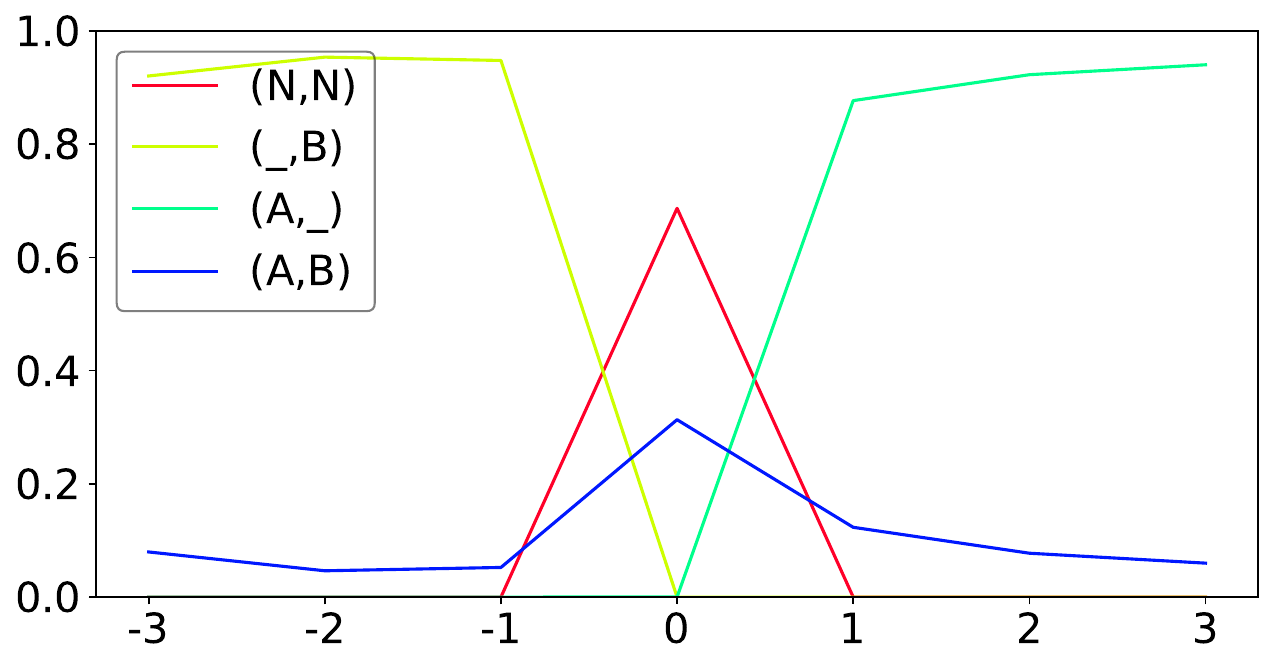} 
		\end{minipage}
		\label{fig:invplp}
	}
    	\subfigure[volatility over last 50 ticks]{
    		\begin{minipage}[b]{0.47\textwidth}
   		 	\includegraphics[width=1\textwidth]{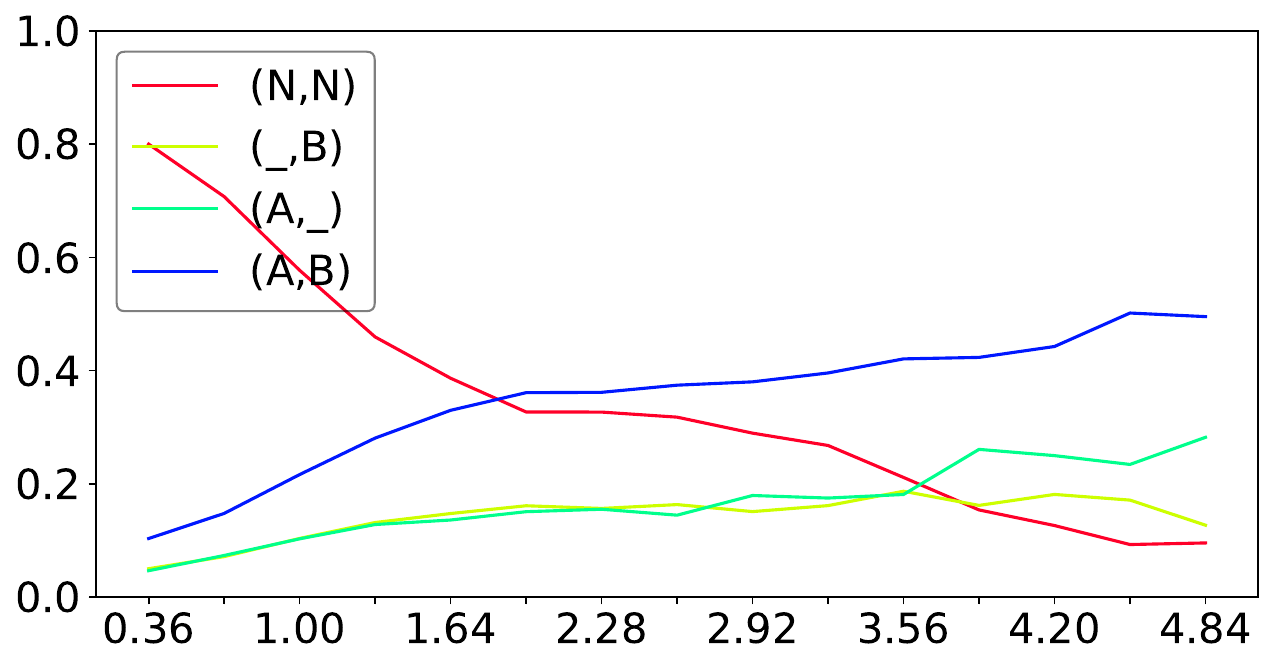}
    		\end{minipage}
		\label{fig:volplp}
    	}
	\\ 
	\subfigure[Spread]{
		\begin{minipage}[b]{0.47\textwidth}
			\includegraphics[width=1\textwidth]{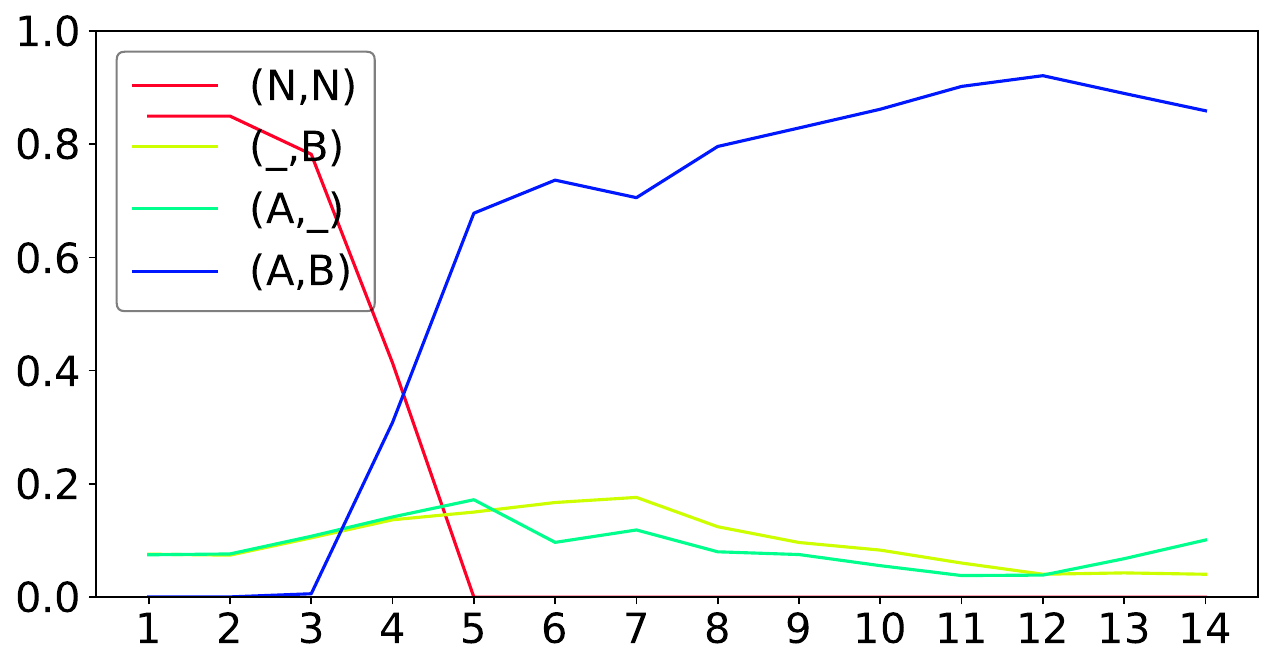} 
		\end{minipage}
		\label{fig:spreadplp}
	}
    	\subfigure[Orderbook Imbalance]{
    		\begin{minipage}[b]{0.47\textwidth}
		 	\includegraphics[width=1\textwidth]{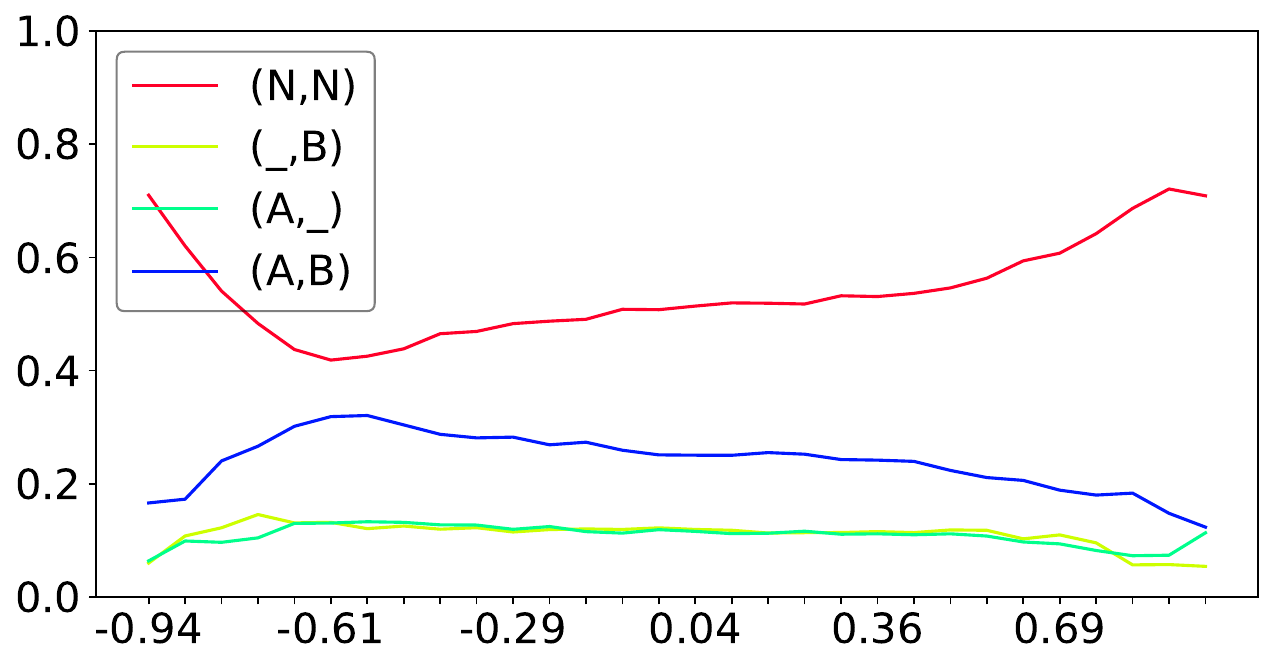}
    		\end{minipage}
		\label{fig:imbpip}
    	}
    \\
	\subfigure[Inventory Value Change]{
		\begin{minipage}[b]{0.47\textwidth}
			\includegraphics[width=1\textwidth]{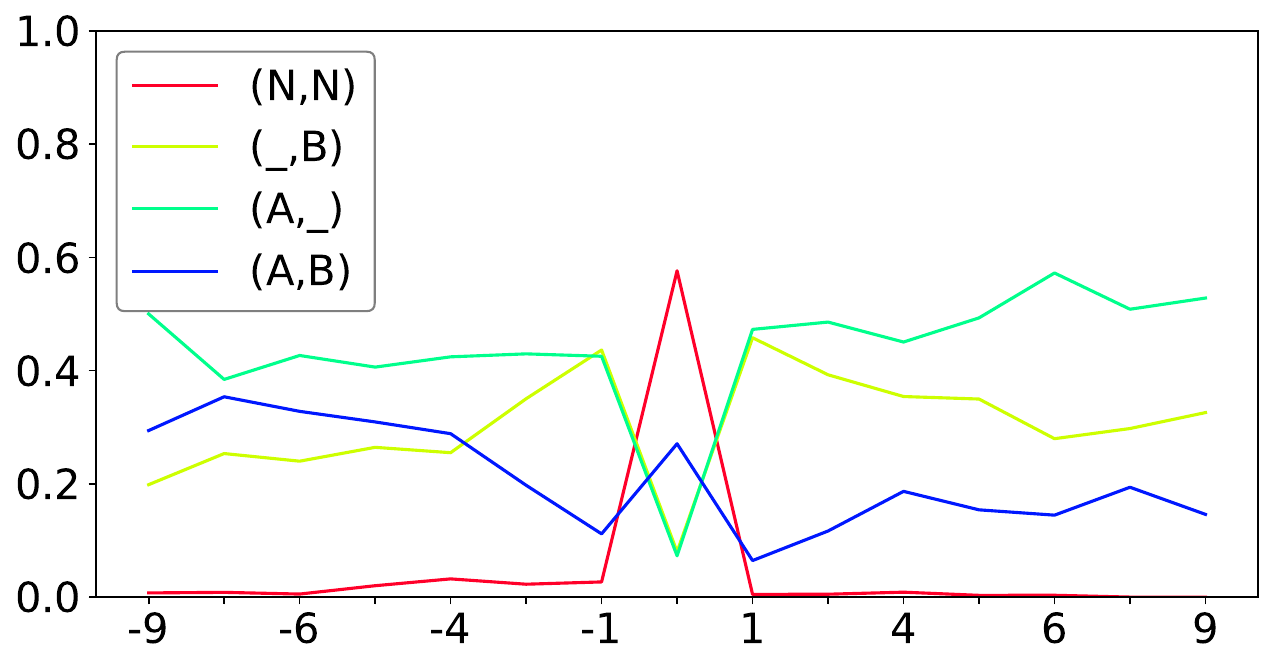} 
		\end{minipage}
		\label{fig:invchangeplp}
	}
	\caption{Different actions in market}
	\label{fig:plp}
\end{figure}

$h_{x}(a)$ is the probability of action $a$ given the state $X=x$. It can be estimated with the datasets $\mathcal{D} $. $X$ refers to the target state, and $Y$ refers to the collection of all other states. Discretization and truncation have been made to guarantee the sample number of $\#N(a,x,\cdot)$ is statistically large enough. For example, when doing the PD analysis for spread, the spread state is truncated between 1 and 15 since the samples with an inventory larger than 15 are rare, and the agent cannot fully learn under those situations. The group of figures \ref{fig:plp} demonstrates the partial dependence relationship. Specifically, we investigate the relationship between actions between inventory level, spread, volatility, orderbook imbalance, and inventory value changes. 

\begin{enumerate}
\item \textbf{Inventory Effect}. For inventory state(Figure \ref{fig:invplp}), the behavior patterns align with traditional market microstructure theories. When the inventory is negative, the agents tend to submit a limited buy order to liquidate the position. It is the same when the inventory is larger than 0. When the inventory is zero, the agent tends to act symmetrically.

\item \textbf{Volatility Effect}. For the volatility state(Figure \ref{fig:volplp}), RL agents tend to make the market when the volatility increases. Maybe it is because when the market is fiercer, the limited order is more likely to be executed.

\item \textbf{Spread Effect}. For the spread state(Figure \ref{fig:spreadplp}), RL agents tend to quote limit order when the spread goes up. When the spread is smaller than 3, the agent will not submit action $(A,B)$ since it is hard to make profits. Since the spread is a potential profit source for the market maker, making the market when the spread is too smaller to cover execution costs is in vain. It sheds like on the market selection. 

\item \textbf{Orderbook Imbalance Effect}. For orderbook imbalance state(Figure \ref{fig:imbpip}), there is inverted U-sharp for $(A,B)$ quote action. When the orderbook Imbalance goes close to 1 or -1, the price movement has a clearer tendency, and the agent is exposed to more adverse selection risks. Therefore, it is wise to reduce the marking market. The inverted U-sharp of symmetric quote action reflects the prudence of the RL agent when there is more adverse selection risk. However, an expected pattern that the agent tends to submit $(A,N)$ more than $(N,B)$ when the imbalance is positive and vice versa is not observed. Also, the lowest point of the U-curve should be at exact 0.5 to reflect a symmetric behavior. There are reasons contributing to those defects. First, the orderbook imbalance is not predictiveness enough, and the model simply overlooks it. Second, the training process for Reinforcement learning is not stable and hinders agents from finding this pattern. To overcome this problem, we suggest eliminating the directional states from $\mathcal{A}$ by introducing Signal Generating Unit\cite{gavsperov2021}. This can not only reduce the state dimension but separate the prediction task so the evaluation and optimization of the predictiveness can be more focused. We leave it to future work.

\item \textbf{Inventory Value Loss Effect}. When the agent suffers little loss from inventory value loss, the agent tends to submit symmetric actions(that is, no action or symmetric quote). 

\end{enumerate}
   
In conclusion, RL agent is inclined to control the inventory and avoid the adverse selection risk by reducing trading when the market moves systematically. Also, It sheds light on the question of which market is more favorable for market making. RL agent tends to make the market when the market is volatile and has a wide spread.

\section{How to be a Good Maker}
\label{cha:exp:MM:adv}

In this section, we summarize the features of being a good market maker. First, one significant trend of the development of market making is the increasingly faster low-latency system. Speed is important in market making. Execution rate decreases in the length of the queue waiting in front of the agent's quote as introduced in the \ref{cha:back:ms}. Therefore, the earlier agent's order can reach the market, the less execution risk it will expose to. Besides, the price process in the high-frequency world is not perfectly efficient, and new information will not be incorporated into the price immediately. The more liquidity the market has, the faster it responds to external shock. Therefore, when diving deep into the micro world, the market becomes increasingly inefficient. Theoretically, with the development of the market and the competition among traders, the market becomes more efficient, and the response time to external shock(News, unexpected large order, etc.) gradually diminish. The relationship between predictiveness and timescale is also observed in China Commodity Market. Therefore, a fast ATS can 1) improve the execution probability and 2) rapidly responds to new information and front-run before others. 

The second point to be a good maker is the adaptation to specific trading rules of the target exchanges. The trading rule differs in exchanges and has a significant influence on trading activity. Trading rules are designed based on the consideration of promoting liquidity and welfare or reducing abnormal volatility and market manipulation. Those trading rules can either be considered in the algorithm or in the design or ATS. For example, The Shanghai Futures Exchange limits the cancel actions number to reduce the ghost liquidity. However, from this study, we found our agent will tend to modify the order frequently. There are two possible solutions to mitigate this problem. The first is modifying Xtrader system to support multi account and implementing a smart order route to bypass this constraint. When a new order is given, it should be distributed to an account with enough cancel action quota. The second solution is incorporating punishment for the cancel action into the reward function so that the agent will take the cost of cancellation into consideration.

Third, We believe incorporating short trend prediction into market making could improve the performance of the agent. For example, incorporating short trend alpha \cite{cartea2018alpha}.


\chapter{Conclusion}
\label{cha:conclusion}

In this study, we investigate the feasibility of applying deep reinforcement learning in market making for the China Commodity Future Market. To our best knowledge, the study of market making via reinforcement learning is relatively rare in China, and this is the first study that applies deep reinforcement learning. This study narrows the gap between the China Commodity Market and a developed market where the market making has been widely studied.

Both model design and trading systems have been studied to achieve this goal. First, we built the Xtrader trading system to handle multi-strategy management and the communication between agents and exchanges. Xtrader achieves inner latency of 400 us. Also, efforts have been invested in the design of RL agents. Comparison experiments with two fixed models demonstrates the RL agent has better control of the inventory and can achieve a better Sharp ratio. Partial dependence analysis(PDA) has been made to analyze the behavior patterns and explain how the agent would react when the environment state varies. It gives our model a sort of explainability to some extent. RL agent demonstrates some interesting behavior patterns in selecting trading timing and trading direction.

There are some points interesting and worth working on further. First of all, continuous effort should be made to further improve the speed of Xtrader. First-tier high-frequency companies achieve nanosecond-level inner latency. Second, market-making is exposed to adverse selection risks. How to incorporate short trend prediction is also worth investigating more. 



\chapter{Supplementary Materirals}
\label{app:A}

\begin{table}
	\centering
	\begin{tabular}{{|c|c|}} \toprule
    \textbf{Parameter name}& \textbf{Parameter Value} \\ \hline
	Learning Rate  & 1e-3    \\ \hline
	Memory Buffer & 50   \\ \hline
	explore Rate & 0.2   \\ \hline
    Training set & 11th Feb - 17th Feb(5 days)  \\ \hline
    validation Set &   18th Feb - 21st Feb(2 days) \\ \hline
    Test Set & 22nd Feb - 28th Feb(5 days) \\ \hline
	Number of entries in dataset & 198,823 \\ \hline
	Inventory Punishment & 0.5  \\  \hline
    Number of Hidden Layer & 3 \\  \hline
    Number of Hidden Unit in Hidden Layer & 16 \\  \hline
	Activation Function for hidden layer & Relu  \\   \hline
    Activation Function for last Layer &  Linear \\   \hline
	evaluation Per Step & 5000 \\ \hline
    Total parameters  &  708  \\   \hline
    Epochs for training  & 30 \\ \hline
	soft update rate & 0.004 \\ \hline
	Optimizer & Adam   \\   \bottomrule
	\end{tabular}
	\caption{Parameters}
	\label{tab:parameters}
\end{table}



\backmatter
\bibliographystyle{plain}
\bibliography{references}

\begin{thebibliography}{10}

\bibitem{abergel2016limitOB}
Fr{\'e}d{\'e}ric Abergel, Marouane Anane, Anirban Chakraborti, Aymen Jedidi,
  and Ioane~Muni Toke.
\newblock {\em Limit order books}.
\newblock Cambridge University Press, 2016.

\bibitem{avellaneda2008high}
Marco Avellaneda and Sasha Stoikov.
\newblock High-frequency trading in a limit order book.
\newblock {\em Quantitative Finance}, 8(3):217--224, 2008.

\bibitem{brogaard2010}
Jonathan Brogaard et~al.
\newblock High frequency trading and its impact on market quality.
\newblock {\em Northwestern University Kellogg School of Management Working
  Paper}, 66, 2010.

\bibitem{budish2015high}
Eric Budish, Peter Cramton, and John Shim.
\newblock The high-frequency trading arms race: Frequent batch auctions as a
  market design response.
\newblock {\em The Quarterly Journal of Economics}, 130(4):1547--1621, 2015.

\bibitem{byrd2019abides}
David Byrd, Maria Hybinette, and Tucker~Hybinette Balch.
\newblock Abides: Towards high-fidelity market simulation for ai research.
\newblock {\em arXiv preprint arXiv:1904.12066}, 2019.

\bibitem{cartea2017algorithmic}
{\'A}lvaro Cartea, Ryan Donnelly, and Sebastian Jaimungal.
\newblock Algorithmic trading with model uncertainty.
\newblock {\em SIAM Journal on Financial Mathematics}, 8(1):635--671, 2017.

\bibitem{cartea2018alpha}
Alvaro Cartea, Sebastian Jaimungal, and Jason Ricci.
\newblock Algorithmic trading, stochastic control, and mutually exciting
  processes.
\newblock {\em SIAM Review}, 60(3):673--703, 2018.

\bibitem{chen2018incorporating}
Yingmei Chen, Zhongyu Wei, and Xuanjing Huang.
\newblock Incorporating corporation relationship via graph convolutional neural
  networks for stock price prediction.
\newblock In {\em Proceedings of the 27th ACM International Conference on
  Information and Knowledge Management}, pages 1655--1658, 2018.

\bibitem{connell1987learning}
Margaret~E Connell and Paul~E Utgoff.
\newblock Learning to control a dynamic physical system.
\newblock {\em Computational intelligence}, 3:330--337, 1987.

\bibitem{dierckx2020using}
Thomas Dierckx, Jesse Davis, and Wim Schoutens.
\newblock Using machine learning and alternative data to predict movements in
  market risk.
\newblock {\em arXiv preprint arXiv:2009.07947}, 2020.

\bibitem{easley2012flow}
David Easley, Marcos~M L{\'o}pez~de Prado, and Maureen O'Hara.
\newblock Flow toxicity and liquidity in a high-frequency world.
\newblock {\em The Review of Financial Studies}, 25(5):1457--1493, 2012.

\bibitem{easley1992adverse}
David Easley and Maureen O'Hara.
\newblock Adverse selection and large trade volume: The implications for market
  efficiency.
\newblock {\em Journal of Financial and Quantitative Analysis}, 27(2):185--208,
  1992.

\bibitem{elainewah2020mmWealfh}
MasonWright ElaineWah and MichaelP Wellman.
\newblock Welfare effects of market making in continuous double auctions:
  Extended abstract.
\newblock {\em cdcdc}, 2018.

\bibitem{figueroa2011estimation}
Jos{\'e}~E Figueroa-L{\'o}pez, Steven~R Lancette, Kiseop Lee, and Yanhui Mi.
\newblock Estimation of nig and vg models for high frequency financial data.
\newblock {\em Handbook of modeling high-frequency data in finance}, 4:3--26,
  2011.

\bibitem{foucault1999order}
Thierry Foucault.
\newblock Order flow composition and trading costs in a dynamic limit order
  market.
\newblock {\em Journal of Financial markets}, 2(2):99--134, 1999.

\bibitem{ganesh2019reinforcement}
Sumitra Ganesh, Nelson Vadori, Mengda Xu, Hua Zheng, Prashant Reddy, and
  Manuela Veloso.
\newblock Reinforcement learning for market making in a multi-agent dealer
  market.
\newblock {\em arXiv preprint arXiv:1911.05892}, 2019.

\bibitem{gao2020optimalLatency}
Xuefeng Gao and Yunhan Wang.
\newblock Optimal market making in the presence of latency.
\newblock {\em Quantitative Finance}, 20(9):1495--1512, 2020.

\bibitem{garman1976market}
Mark~B Garman.
\newblock Market microstructure.
\newblock {\em Journal of financial Economics}, 3(3):257--275, 1976.

\bibitem{gavsperov2021review}
Bruno Ga{\v{s}}perov, Stjepan Begu{\v{s}}i{\'c}, Petra
  Posedel~{\v{S}}imovi{\'c}, and Zvonko Kostanj{\v{c}}ar.
\newblock Reinforcement learning approaches to optimal market making.
\newblock {\em Mathematics}, 9(21):2689, 2021.

\bibitem{gavsperov2021reinforcement}
Bruno Ga{\v{s}}perov, Stjepan Begu{\v{s}}i{\'c}, Petra
  Posedel~{\v{S}}imovi{\'c}, and Zvonko Kostanj{\v{c}}ar.
\newblock Reinforcement learning approaches to optimal market making.
\newblock {\em Mathematics}, 9(21):2689, 2021.

\bibitem{gavsperov2021}
Bruno Ga{\v{s}}perov and Zvonko Kostanj{\v{c}}ar.
\newblock Market making with signals through deep reinforcement learning.
\newblock {\em IEEE Access}, 9:61611--61622, 2021.

\bibitem{gastli2021satelline}
Mohamed~Sadok Gastli.
\newblock Deep learning tools for yield and price forecasting using satellite
  images.
\newblock Master's thesis, University of Waterloo, 2021.

\bibitem{glosten1985bid}
Lawrence~R Glosten and Paul~R Milgrom.
\newblock Bid, ask and transaction prices in a specialist market with
  heterogeneously informed traders.
\newblock {\em Journal of financial economics}, 14(1):71--100, 1985.

\bibitem{grossman1988}
Sanford~J Grossman.
\newblock Program trading and market volatility: A report on interday
  relationships.
\newblock {\em Financial Analysts Journal}, 44(4):18--28, 1988.

\bibitem{gueant2019deep}
Olivier Gu{\'e}ant and Iuliia Manziuk.
\newblock Deep reinforcement learning for market making in corporate bonds:
  beating the curse of dimensionality.
\newblock {\em Applied Mathematical Finance}, 26(5):387--452, 2019.

\bibitem{pham2013}
Fabien Guilbaud and Huyen Pham.
\newblock Optimal high-frequency trading with limit and market orders.
\newblock {\em Quantitative Finance}, 13(1):79--94, 2013.

\bibitem{haider2019effect}
Abbas Haider, Hui Wang, Bryan Scotney, and Glenn Hawe.
\newblock Effect of market spread over reinforcement learning based market
  maker.
\newblock In {\em International Conference on Machine Learning, Optimization,
  and Data Science}, pages 143--153. Springer, 2019.

\bibitem{hansch1998inv}
Oliver Hansch and Naik.
\newblock Do inventories matter in dealership markets, evidence from the london
  stock exchange.
\newblock {\em The Journal of Finance}, 53(5):1623--1656, 1998.

\bibitem{ho1981optimal}
Thomas Ho and Hans~R Stoll.
\newblock Optimal dealer pricing under transactions and return uncertainty.
\newblock {\em Journal of Financial economics}, 9(1):47--73, 1981.

\bibitem{Huang2019ats}
Boming Huang, Yuxiang Huan, Li~Da Xu, Lirong Zheng, and Zhuo Zou.
\newblock Automated trading systems statistical and machine learning methods
  and hardware implementation: a survey.
\newblock {\em Enterprise Information Systems}, 13(1):132--144, 2019.

\bibitem{huang1997components}
Roger~D Huang and Hans~R Stoll.
\newblock The components of the bid-ask spread: A general approach.
\newblock {\em The Review of Financial Studies}, 10(4):995--1034, 1997.

\bibitem{karpe2020multi}
Mich{\"a}el Karpe, Jin Fang, Zhongyao Ma, and Chen Wang.
\newblock Multi-agent reinforcement learning in a realistic limit order book
  market simulation.
\newblock In {\em Proceedings of the First ACM International Conference on AI
  in Finance}, pages 1--7, 2020.

\bibitem{kercheval2011risk}
Alec~N Kercheval and Yang Liu.
\newblock Risk forecasting with garch, skewed t distributions, and multiple
  timescales.
\newblock {\em Handbook of Modeling High-Frequency Data in Finance}, 4:163,
  2011.

\bibitem{konidaris2011value}
George Konidaris, Sarah Osentoski, and Philip Thomas.
\newblock Value function approximation in reinforcement learning using the
  fourier basis.
\newblock In {\em Twenty-fifth AAAI conference on artificial intelligence},
  2011.

\bibitem{korajczyk2019high}
Robert~A Korajczyk and Dermot Murphy.
\newblock High-frequency market making to large institutional trades.
\newblock {\em The Review of Financial Studies}, 32(3):1034--1067, 2019.

\bibitem{kyle1985continuous}
Albert~S Kyle.
\newblock Continuous auctions and insider trading.
\newblock {\em Econometrica: Journal of the Econometric Society}, pages
  1315--1335, 1985.

\bibitem{li2020generating}
Junyi Li, Xintong Wang, Yaoyang Lin, Arunesh Sinha, and Michael Wellman.
\newblock Generating realistic stock market order streams.
\newblock In {\em Proceedings of the AAAI Conference on Artificial
  Intelligence}, volume~34, pages 727--734, 2020.

\bibitem{mani2019applications}
Mohammad Mani, Steve Phelps, and Simon Parsons.
\newblock Applications of reinforcement learning in automated market-making.
\newblock In {\em Proceedings of the GAIW: Games, Agents and Incentives
  Workshops, Montreal, Canada}, pages 13--14, 2019.

\bibitem{merton1975optimum}
Robert~C Merton.
\newblock Optimum consumption and portfolio rules in a continuous-time model.
\newblock In {\em Stochastic optimization models in finance}, pages 621--661.
  Elsevier, 1975.

\bibitem{mnih2015human}
Volodymyr Mnih, Koray Kavukcuoglu, David Silver, Andrei~A Rusu, Joel Veness,
  Marc~G Bellemare, Alex Graves, Martin Riedmiller, Andreas~K Fidjeland, Georg
  Ostrovski, et~al.
\newblock Human-level control through deep reinforcement learning.
\newblock {\em nature}, 518(7540):529--533, 2015.

\bibitem{morton1977discounting}
Thomas~E Morton and William~E Wecker.
\newblock Discounting, ergodicity and convergence for markov decision
  processes.
\newblock {\em Management Science}, 23(8):890--900, 1977.

\bibitem{parlour1998price}
Christine~A Parlour.
\newblock Price dynamics in limit order markets.
\newblock {\em The Review of Financial Studies}, 11(4):789--816, 1998.

\bibitem{pithyachariyakul1986exchange}
Pipat Pithyachariyakul.
\newblock Exchange markets: a welfare comparison of market maker and walrasian
  systems.
\newblock {\em The Quarterly Journal of Economics}, 101(1):69--84, 1986.

\bibitem{rummery1994line}
Gavin~A Rummery and Mahesan Niranjan.
\newblock {\em On-line Q-learning using connectionist systems}, volume~37.
\newblock Citeseer, 1994.

\bibitem{schaul2016prioritized}
Tom Schaul, John Quan, Ioannis Antonoglou, and David Silver.
\newblock Prioritized experience replay.
\newblock In {\em ICLR (Poster)}, 2016.

\bibitem{schumaker2009textual}
Robert~P Schumaker and Hsinchun Chen.
\newblock Textual analysis of stock market prediction using breaking financial
  news: The azfin text system.
\newblock {\em ACM Transactions on Information Systems (TOIS)}, 27(2):1--19,
  2009.

\bibitem{shanghai2021exchange}
SHFE.
\newblock Articles of association of shanghai futures exchange.
\newblock \url{http://www.shfe.com.cn/regulation/regulation/rules/}, May 2022.

\bibitem{spooner2018}
Thomas Spooner, John Fearnley, Rahul Savani, and Andreas Koukorinis.
\newblock Market making via reinforcement learning.
\newblock {\em arXiv preprint arXiv:1804.04216}, 2018.

\bibitem{spooner2020robust}
Thomas Spooner and Rahul Savani.
\newblock Robust market making via adversarial reinforcement learning.
\newblock {\em arXiv preprint arXiv:2003.01820}, 2020.

\bibitem{stoikov2009option}
Sasha Stoikov and Mehmet Sa{\u{g}}lam.
\newblock Option market making under inventory risk.
\newblock {\em Review of Derivatives Research}, 12(1):55--79, 2009.

\bibitem{stoll1978supply}
Hans~R Stoll.
\newblock The supply of dealer services in securities markets.
\newblock {\em The Journal of Finance}, 33(4):1133--1151, 1978.

\bibitem{subramoni2010streaming}
Hari Subramoni, Fabrizio Petrini, Virat Agarwal, and Davide Pasetto.
\newblock Streaming, low-latency communication in on-line trading systems.
\newblock In {\em 2010 IEEE International Symposium on Parallel \& Distributed
  Processing, Workshops and Phd Forum (IPDPSW)}, pages 1--8. IEEE, 2010.

\bibitem{tang2016fpga}
Qiu Tang, Majing Su, Lei Jiang, Jiajia Yang, and Xu~Bai.
\newblock A scalable architecture for low-latency market-data processing on
  fpga.
\newblock In {\em 2016 IEEE Symposium on Computers and Communication (ISCC)},
  pages 597--603. IEEE, 2016.

\bibitem{thung2018brief}
Kim-Han Thung and Chong-Yaw Wee.
\newblock A brief review on multi-task learning.
\newblock {\em Multimedia Tools and Applications}, 77(22):29705--29725, 2018.

\bibitem{van2016deep}
Hado Van~Hasselt, Arthur Guez, and David Silver.
\newblock Deep reinforcement learning with double q-learning.
\newblock In {\em Proceedings of the AAAI conference on artificial
  intelligence}, volume~30, 2016.

\bibitem{vyetrenko2020get}
Svitlana Vyetrenko, David Byrd, Nick Petosa, Mahmoud Mahfouz, Danial Dervovic,
  Manuela Veloso, and Tucker Balch.
\newblock Get real: Realism metrics for robust limit order book market
  simulations.
\newblock In {\em Proceedings of the First ACM International Conference on AI
  in Finance}, pages 1--8, 2020.

\bibitem{watkins1992q}
Christopher~JCH Watkins and Peter Dayan.
\newblock Q-learning.
\newblock {\em Machine learning}, 8(3):279--292, 1992.

\bibitem{watkins1989learning}
CJCH Watkins.
\newblock Learning from delayed rewards.
\newblock {\em PhD thesis, King's College, University of Cambridge}, 1989.

\bibitem{zhang2020reinforcement}
Ge~Zhang and Ying Chen.
\newblock Reinforcement learning for optimal market making with the presence of
  rebate.
\newblock {\em Available at SSRN 3646753}, 2020.

\bibitem{zhong2020}
Yueyang Zhong, YeeMan Bergstrom, and Amy Ward.
\newblock Data-driven market-making via model-free learning.
\newblock In {\em IJCAI}, pages 4461--4468, 2020.

\end{thebibliography}

\end{document}